\documentclass[twocolumn]{biophys-new}
\usepackage[utf8]{inputenc}
\usepackage{graphicx}
\usepackage{xr-hyper}
\usepackage[colorlinks,allcolors=cyan!70!black]{hyperref}
\usepackage{amsfonts}
\def\pombe{\textit{S.~pombe}}

\def\e{\mathbf{e}}

\def\r{\mathbf{r}}
\def\x{\mathbf{x}}

\def\X{\mathbf{X}}

\def\vecU{\mathbf{U}}
\def\U{\mathbf{U}}

\def\vtheta{\boldsymbol{\theta}}

\def\F{\mathbf{F}}

\def\torque{\mathbf{T}}

\newcommand{\dv}[2][]{\frac{\mathrm{d}#1}{\mathrm{d}#2}}

\newcommand{\dd}[2][]{\hspace{.025\baselineskip}\mathrm{d}^{#1} #2 \hspace{.075\baselineskip}}
\def\lbo{\Delta}
\def\cotlap{\Delta^{\text{cot}}}

\newcommand{\texpect}[1]{\left\langle  #1 \right\rangle}
\newcommand{\saverage}[1]{\overline{#1}}

\title{Dynamics of the Nuclear Envelope and Mitotic Spindle in Closed Mitosis}
\runningtitle{Dynamics of the Nuclear Envelope and Mitotic Spindle in Closed Mitosis} 

\author[1]{Wilson Lough}
\author[1,2]{Zachary R Gergely}
\author[3]{Eduardo Corona}
\author[1,2,*]{Meredith D Betterton}
\runningauthor{Lough \textit{et al}.} 

\affil[1]{Department of Physics, University of Colorado, Boulder, Colorado}
\affil[2]{Department of Molecular, Cellular, and Developmental Biology, University of Colorado, Boulder, Colorado}
\affil[3]{Department of Applied Mathematics, University of Colorado, Boulder, Colorado}

\corrauthor[*]{mdb@colorado.edu}

\papertype{Article}

\begin{document}

\begin{frontmatter}




\begin{abstract}
  During closed mitosis, large-scale deformation of the nuclear
  envelope is driven by the mitotic spindle and chromosomes contained
  within it. Previous models of closed mitosis have established that
  the spatial distribution of spindle-generated forces influences
  nuclear morphology, but have imposed axisymmetry and focused
  primarily on equilibrium shapes. We introduce a three-dimensional
  biophysical model of nuclear envelope remodeling and spindle
  elongation that incorporates thermal fluctuations to examine how
  spindle-generated forces, chromatin distribution, and nuclear
  envelope elasticity control nuclear shape
  in fission yeast. The model reproduces features of both successful
  and aberrant nuclear division, including centering and alignment of
  initially mispositioned spindles, the spherical-to-dumbbell nuclear
  shape transition, and the formation of thin nuclear
  protrusions. Nuclear morphology is governed primarily by two
  dimensionless quantities: the ratio of maximum spindle-generated
  force to nuclear envelope bending force and the size of
  spindle-associated chromatin clusters in contact with the envelope
  relative to the overall size of the nucleus. A phase diagram of
  nuclear shape shows that insufficient spindle force produces
  elongated but undivided nuclei, while sufficiently large spindle
  forces produce either symmetric division or envelope tethers,
  depending on the size of the chromatin-envelope contact region. Near
  the boundary between symmetric division and tethering, nuclei
  develop asymmetric single-tether configurations. A
  reduced-order representation of the coupled envelope-spindle system
  shows a symmetry-breaking transition between symmetric and
  asymmetric nuclear shapes. Below a critical
  chromatin-envelope contact size, symmetric configurations lose
  stability during spindle elongation and thermal fluctuations drive
  nuclear shape trajectories toward one of two stable asymmetric
  configurations. These findings identify a growth-driven mechanical
  mechanism for spindle centering and alignment during early mitosis
  and reveal how asymmetric nuclear morphologies can arise during
  anaphase even in the absence of structural differences between the
  two spindle poles.
\end{abstract}

\begin{sigstatement}
In closed mitosis, the nucleus divides with the nuclear envelope
intact. In fission yeast, the spindle stretches the envelope into two
equal daughter nuclei, and failures produce misshapen nuclei or thin
envelope tethers. Previous biophysical models assumed axial symmetry
or solved for equilibrium shapes, so they could not follow dynamics of
the spindle and envelope. We built a three-dimensional dynamic model
of the envelope, spindle, and chromatin including thermal
fluctuations. Symmetric division requires spindle force to be spread
over envelope regions much larger than the spindle pole bodies. When
the chromatin contact region is too small, fluctuations drive nuclei
into asymmetric states. These results link chromatin organization at
the spindle poles to the fidelity of nuclear division.
\end{sigstatement}
\end{frontmatter}

\begin{figure*}[t]
    \centering
    \includegraphics[width=6.75in]{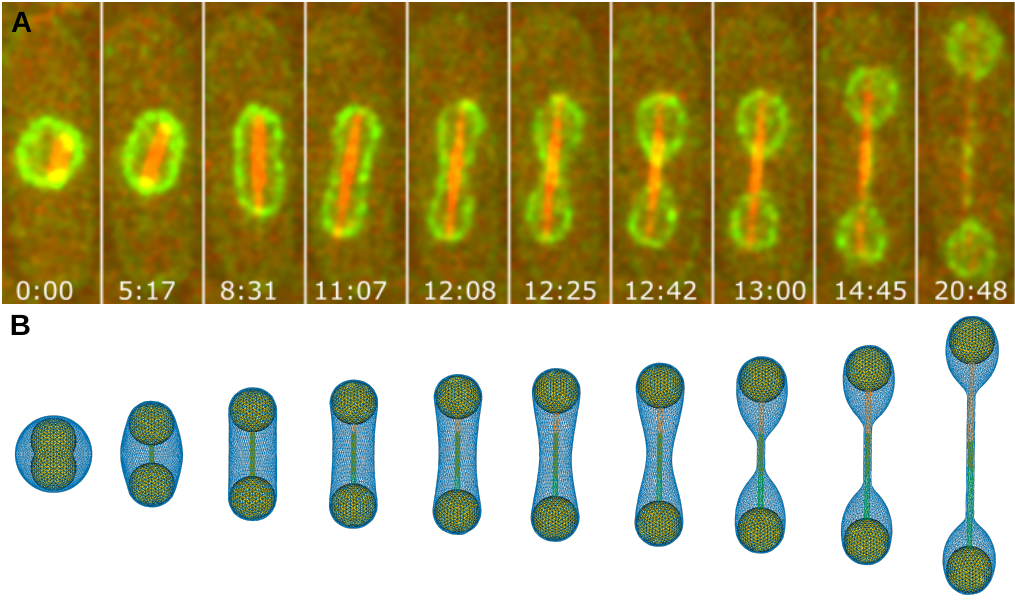}
    \caption{Nuclear envelope shape transitions during closed mitosis
    in fission yeast. (A) Nuclear envelope (green, Cut11-GFP) and
    spindle microtubules (red, mCherry-atb2) from confocal
    fluorescence microscopy shown as maximum-intensity-projected image
    sequence. Time in min:sec. (B) Corresponding sequence of simulated nuclear shape transitions with nuclear envelope shown in blue,
    SPB-chromatin complexes in yellow.}
      \label{fig:mitosis_sim_timelapse}
\end{figure*}

\section*{Introduction}

Successful mitosis requires coordinated remodeling and reorganization
of cellular structures~\cite{mcintosh16}. One of the less
well-understood remodeling processes is that of the nuclear envelope
(NE), a double-membrane system found in eukaryotes which separates
genetic material in the nucleus from the rest of the
cell~\cite{zhang13, dey21}.  Strategies for the segregation and
movement of chromosomes during mitosis differ in the degree of NE
breakdown~\cite{boettcher13, smoyer14}. At the two extremes of this
continuum are open mitosis, in which the NE is disassembled and later
reassembled in each of the two daughter cells, and closed mitosis, in
which the NE remains largely intact and must undergo large
deformations to accommodate chromosome segregation. During closed
mitosis in the fission yeast \textit{Schizosaccharomyces pombe}, this
occurs via changes of NE shape from a spherical to a dumbbell-like
configuration (Fig.~\ref{fig:mitosis_sim_timelapse}) with local
disassembly confined to a narrow neck connecting the pair of
proto-daughter nuclei~\cite{exposito-serrano20, dey20}.

The NE is made up of two concentric lipid bilayers, the inner and
outer nuclear membranes. These two membranes are fused at nuclear
pores, regions of high curvature which are supported by ring-like
protein assemblies called nuclear pore complexes (NPCs)~\cite{maul77,
  torbati16a, hamed21, noguchi16}. The mechanical properties of the
NE depend on its lipid composition and the membrane-associated
proteins it contains~\cite{agrawal19, gonzalez12,
  romanauska23}. Because yeast lack a nuclear
lamina~\cite{gonzalez12}, the mechanics of the nuclear membranes play
a more prominent role in determining nuclear morphology than in animal
cells. Despite its complex microscopic structure, previous work has
shown that the large-scale mechanical behavior of the NE is well
described by the Helfrich model commonly used for single lipid
bilayers~\cite{lim07}.

In preparation for mitosis, interphase chromatin condenses to form
mitotic chromosomes. The degree of condensation varies between
different organisms, and these differences in chromatin structure
change nuclear mechanics~\cite{miroshnikova22, stephens18,
  schreiner15, guilak00, dahl05, pajerowski07, wintner20, stephens17,
  chalut12, furusawa15, stephens19a}. Fluorescence imaging in \pombe\
has found that mitotic chromatin is more diffuse than the compact
structures seen in animal cells~\cite{west02,
  mora-bermudez07}. Chromosome volume decreases by only 20\% during
mitosis in \pombe~\cite{gerguri21}, compared to a factor of 2-3
decrease observed in mammalian cells~\cite{mora-bermudez07}.
The relatively diffuse mitotic chromatin in \pombe\ suggests that
chromatin clustered at the spindle poles may occupy a substantial
fraction of the nuclear volume and distribute spindle-generated forces
over regions of the envelope much larger than the spindle-pole bodies (SPBs)
themselves~\cite{zheng07, castagnetti15, zhu16}.

Segregation of chromosomes is orchestrated by the mitotic spindle, an
assembly of microtubules, molecular motors, and cross-linking and
associated proteins \cite{mcintosh16, elting18, pavin16}. Formation of
the spindle in fission yeast begins with the nucleation of
microtubules at the SPBs, a pair of specialized microtubule organizing
centers embedded in the NE \cite{ding93, ding97, tallada09}. As the
bipolar spindle forms, SPBs move towards opposite sides of the nucleus
while interpolar microtubules form antiparallel overlaps and
kinetochore microtubules capture and orient chromosomes \cite{hagan92,
  ding93, loiodice05, kalinina12, gay12, blackwell17a, edelmaier20,
  gergely23b, jones24}.
As the cell enters anaphase, sister chromatids separate and are pulled
towards opposite spindle poles by shortening microtubules
\cite{funabiki93, grishchuk06, betterton13}. Segregated chromatids
cluster near the two spindle poles and the inner nuclear membrane
\cite{buttrick11}. In
anaphase B, motor proteins slide apart overlapping antiparallel
microtubules, driving spindle elongation and further separating sister
chromatids \cite{mallavarapu99, tolic-norrelykke04, loiodice05, fu09,
  kruger19, gergely23, gergely23b}. Extensile forces produced within
the spindle are transmitted to the NE and drive a sequence of nuclear
shape transitions which ultimately result in the formation of separate
daughter nuclei (Fig.~\ref{fig:mitosis_sim_timelapse}) \cite{zheng07,
  zhu16, dey20, exposito-serrano20, begley25, real-calderon26,
  mahoney26}.

Mechanical interactions between the NE, spindle, and chromosomes are
important to coordinate spindle elongation and positioning with NE
remodeling \cite{zheng07, yam11, begley25}. Recent experiments in
fission yeast have demonstrated that the spindle and NE exert forces
that shape one another \cite{begley25, real-calderon26,
  mahoney26, zhou26a}. Aberrations in spindle structure, chromatin
localization, or NE mechanics disrupt this coordination and perturb
nuclear division. Disruption of spindle pole structure or failure of
chromosomes to stably attach to the spindle can cause the nucleus to
become misshapen or form thin NE protrusions \cite{zheng07, toya07,
  castagnetti15, yukawa15, zhu16, syrovatkina15, gergely16,
  gergely23b, jones24, zhou26a}. Suppression of lipid synthesis,
which restricts the NE area increase that normally occurs during
closed mitosis, leads to
excessive force that can buckle or break the spindle \cite{saitoh96,
  yam11, takemoto16, makarova16, begley25}. Excess area increase due
to over-synthesis of membrane lipids can result in misshapen nuclei
\cite{tange02, makarova16, sajiki18, kume19, dey20}.

Previous models of mitosis have investigated the capture of
chromosomes \cite{hill85, holy94, wollman05, paul09, kalinina12,
  blackwell17}, regulation of spindle length \cite{goshima05,
  brust-mascher04, mogilner06, edelmaier20}, control of spindle
position \cite{dogterom98, grill05, kozlowski07}, or aspects of
spindle assembly \cite{nedelec02, cytrynbaum03, schaffner06,
  burbank07, loughlin10}. Detailed models of spindle assembly and
chromosome segregation \cite{blackwell17a, rincon17, lamson19,
  edelmaier20} are able to quantitatively match experimental
measurements of spindle assembly timing, length, and structure during
early mitosis \cite{edelmaier20, blackwell17a}.  However, they assume
a spherical nucleus and so cannot capture the NE dynamics in anaphase
and nuclear division. There has also been work which combined NE
mechanics with minimal models of microtubule bundles, the spindle, or
spindle assembly \cite{lim07, lim09, castagnetti15, zhu16, fang20,
  begley25, zhou26a}. This work has shown that tethers form when force
from a microtubule bundle or the spindle is applied over a small
region of the envelope \cite{lim07, zhu16, zhou26a}, as expected from
the theory of membrane tube formation \cite{derenyi02}. Chromosomes
clustered at the spindle poles favor symmetric dumbbell shapes
\cite{castagnetti15}, and NE tension compresses the spindle and can
cause it to buckle \cite{begley25}. Previous models of NE shape imposed
axisymmetry. Most compute equilibrium shapes by energy minimization at
a set length or force \cite{lim07, lim09, castagnetti15, zhu16}. The
one dynamic model coupled NE shape change to a mean-field description
of microtubule growth and motor-driven sliding \cite{fang20}. Some
models prescribed the distribution of spindle force on the NE rather
than computing it from chromatin contact \cite{zhu16, fang20}. How
mechanical interactions between the spindle and NE regulate spindle
positioning and the stability of symmetric nuclear configurations
during spindle elongation remains poorly understood.

Here we develop a mechanical model of the fission yeast nucleus during
closed mitosis that describes the three-dimensional stochastic
dynamics of the NE and mitotic spindle, accounting for large-scale
envelope deformation, spindle elongation, and force transmission
through SPB-chromatin complexes. The model couples a dynamically
triangulated Helfrich description of the NE~\cite{helfrich73} with a
coarse-grained mechanical description of the mitotic spindle and
chromatin, incorporating thermal fluctuations of the NE without
imposing axisymmetry.  We examine how force transmission through the
spindle-chromatin-NE system controls nuclear shape change and spindle
positioning during closed mitosis.  Interactions between the
elongating spindle and deformable NE robustly center and align
initially mispositioned spindles.  Nuclear morphology depends on the
magnitude of spindle-generated force and the size of chromatin
clusters in contact with the NE. We identify distinct mechanical
regimes corresponding to elongated but undivided nuclei, symmetric
division, and the formation of nuclear tethers. Finally, we introduce
reduced-order shape descriptors for the NE-spindle system and show
that variations in the degree of chromatin-NE contact can drive
symmetry-breaking transitions between morphologies with or without
reflection symmetry about the spindle midplane. 

\section*{Methods}

\begin{figure}
    \centering
    \includegraphics[width=3.25in]{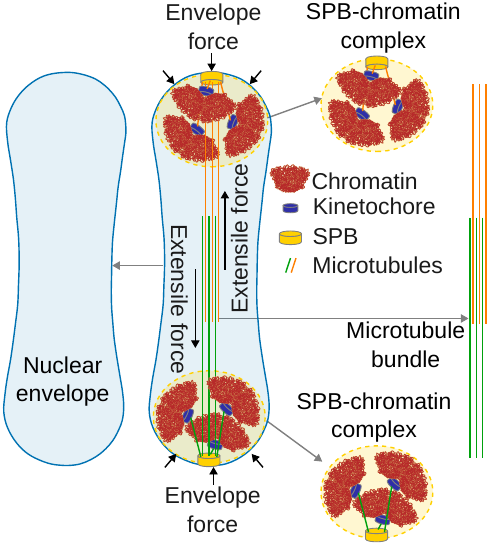}
    \caption{\small 
    Mechanical model of the fission yeast nucleus during closed mitosis. The model consists of a deformable NE, an extensible microtubule bundle representing the mitotic spindle, and two SPB-chromatin complexes which transmit spindle-generated extensile forces to the NE. 
      }
    \label{fig:nucleus_diagram}
\end{figure}

Our model has three components (Fig.~\ref{fig:nucleus_diagram}): a
deformable NE, an extensible microtubule bundle representing the
spindle, and two SPB-chromatin complexes that transmit force from the
bundle to the NE. The dynamics of the system are governed by
overdamped Langevin equations. Here we summarize the model; for
details, see the supplementary material.

\subsection*{Nuclear envelope}

Because fission yeast nuclei lack lamins, vesicle-like membrane
mechanics capture features of nuclear shape \cite{lim07, yam11}. We
therefore model the NE as a two-dimensional fluid with bending
elasticity and surface tension. Bending elasticity is described by the
Helfrich energy~\cite{helfrich73}
\begin{gather}
\label{eq:envelope_energy}
U_{\text{bend}}=\int_{M} 2BH^2\dd{A},
\end{gather}
where $M$ is the envelope surface, $B$ is the bending modulus, $H$ is
the mean curvature, and we have set spontaneous curvature to zero.
During mitosis in \pombe, membrane moves from the endoplasmic
reticulum to the NE and the NE area increases. Surface
tension therefore has a spatially uniform part $\gamma_0$, associated
with lipid availability, and a position dependent part that increases
linearly with local area strain \cite{deserno15}. Because enclosed
nuclear volume $V$ remains approximately conserved during mitosis in
\pombe~\cite{lim07, yam11}, we enforce volume conservation using a
 constraint potential
\begin{gather}
U_{\text{vol}}=\frac{k_{\text{vol}}}{2V_0}(V-V_0)^2,
\end{gather}
where $V_0$ is the reference volume and $k_{\text{vol}}$ the volume
expansion modulus. We neglect hydrodynamic coupling between different
points on the NE and approximate viscous interactions with the
surrounding fluid using a local drag coefficient $\zeta$ per unit
area. Thermal fluctuations are represented by a Gaussian stochastic
traction whose covariance is determined by the fluctuation-dissipation
theorem.

To discretize the NE, we approximate $M$ by a piecewise linear
triangulated surface \cite{vutukuri20}. Each vertex is assigned a dual cell constructed
by connecting the centroids of incident triangles and the midpoints of
incident edges. Tractions are integrated over dual cells to obtain
discrete vertex forces. The resulting discrete viscous force
effectively assigns each vertex an isotropic drag tensor
$\boldsymbol{\Gamma}_i=\zeta a_i^*\mathbf{I}$, where $a_i^*$ is the
area of vertex $i$'s dual cell. We approximate the surface tension
field as piecewise constant, assigning each triangular face $f$ a
tension
\begin{gather}
\label{eq:NE_face_tension}
\gamma_f=\gamma_0+\frac{k_{\text{area}}}{a_0}(a_f-a_0),
\end{gather}
where $a_f$ is the face area, $a_0$ the mean initial face area, and
$k_{\text{area}}$ the local area expansion modulus. We also introduce
an edge-length tethering potential and edge flips to regulate mesh
geometry and control the effective two-dimensional viscosity of the
envelope~\cite{noguchi05}.

The NE interacts with the spindle through steric repulsion. NE
vertices repel the vertices of the SPB-chromatin complexes and the
surface of the microtubule bundle through the Weeks-Chandler-Andersen
(WCA) potential \cite{weeks71}
\begin{gather}
\label{eq:wca_potential}
    U_{\text{WCA}}(d)
    =
    \begin{cases}
    4\varepsilon
    \left[
    \left(\frac{\sigma}{d}\right)^{12}
    -
    \left(\frac{\sigma}{d}\right)^6
    \right]
    +
    \varepsilon & d\leq 2^{1/6} \sigma\\
    0 & d>2^{1/6} \sigma
    \end{cases}
\end{gather}
where $d$ is the separation, $\varepsilon$ sets the strength, and
$\sigma$ sets the range. Vertex dynamics are governed by a coupled
system of overdamped Langevin equations
\begin{gather}
\label{eq:vertex_langevin}
\boldsymbol{\Gamma}_i\cdot\mathbf{v}_i
=
\F_{i}
+
\frac{k_BT}{3}(\cotlap\x)_i
+
\sqrt{2k_BT}\boldsymbol{\Gamma}_i^{1/2}\cdot\boldsymbol{\xi}_i,
\end{gather}
where $\F_i$ is the sum of bending, tension, pressure, tethering, and
contact forces, and $\boldsymbol{\xi}_i$ are independent
three-dimensional Gaussian white noise processes. Because vertex drag
tensors depend on the instantaneous NE geometry, the noise is
multiplicative. The term $k_BT(\cotlap\x)_i/3$ is the thermal drift
that thermodynamic consistency requires in the It\^o interpretation
\cite{lau07, delong14}.

\subsection*{Spindle}

The spindle is modeled as a bundle of two partially overlapping
antiparallel half-bundles
(Fig. \ref{S-fig:mt_bundle_config_and_velocities}). Each half-bundle
is a rigid cylinder of fixed radius $a$ whose length $L_{\text{half}}$ grows at a
prescribed speed $v_{\text{poly}}$ \cite{betterton13}. The
half-bundles slide relative to one another along their shared axis,
which changes the total length $L$ and the overlap length $\ell=2L_{\text{half}}-L$.
Motors in the overlap exert an extensile force density $f_{\text{motor}}$ \cite{fu09,
  kruger19, gergely23b, jones24}, so the total sliding force is
$\ell f_{\text{motor}}$. Growth stops when $\ell f_{\text{motor}}$ reaches a maximum
$F_{\text{max}}$, which sets the maximum force the spindle can exert.

The bundle's independent degrees of freedom are described by its total
length $L$, midpoint position $\X_{\text{mt}}$, and spatial orientation,
represented by an orthogonal matrix $\U_{\text{mt}}$. Viscous drag on the
half-bundles is modeled using local slender-body theory \cite{cox70}.
The translational and rotational dynamics of the bundle are governed
by overdamped Langevin equations describing force and torque balance
that include interactions with the SPB-chromatin complexes and NE,
viscous drag, and thermal fluctuations. Relative sliding of the
half-bundles is deterministic and obeys
\begin{gather}
\label{eq:mt_sliding}
\Gamma_L v_L = Q_L^{\text{spb}} + \ell f_{\text{motor}},
\\
\label{eq:mt_overlap_length_evolution}
\dv[\ell]{t}=2v_{\text{poly}}-v_L,
\end{gather}
where $v_L=\dv[L]{t}$ is the sliding velocity, $\Gamma_L$ the sliding
drag, and $Q_L^{\text{spb}}$ the generalized force conjugate to $L$
due to interactions with SPB-chromatin complexes. NE contact acts on
the sides of the bundle, normal to its axis, and does not contribute
to sliding.

\subsection*{SPB-chromatin complexes}

At each spindle pole, spindle force is transmitted to the NE through
the SPB and chromatin clustered near it. We treat each SPB-chromatin
complex as a rigid sphere of radius $r_{\text{complex}}$. Early in
mitosis $r_{\text{complex}}$ is approximately the SPB radius. After
chromatin has segregated to the poles, $r_{\text{complex}}$ can be a
sizable fraction of the nuclear radius $R$. The surface of each
complex is triangulated, and its vertices interact with NE vertices
through the WCA potential (Eq. \ref{eq:wca_potential}). A linear
spring and a torsional spring with $\F^{\text{mt}}=-k_{\text{spb}}\Delta\X_{\text{spb}}$ and
$\torque^{\text{mt}}=-k_{\text{rot}}\Delta\vtheta_{\text{spb}}$ hold each complex
at one end of the bundle and aligned with it (Fig.
\ref{S-fig:mt_spb_interactions}). The complexes have Stokes
translational and rotational drag.

\subsection*{Numerical integration}

We discretize the overdamped Langevin equations describing force and
torque balance using the standard Euler-Maruyama scheme for
translational degrees of freedom~\cite{higham01} and the exponential
Euler-Maruyama scheme for rotational degrees of
freedom~\cite{piggott16}. For rotational motion of the SPB-chromatin
complexes and microtubule bundle, multiplicative updates preserve the
orthonormality of orientation matrices. For NE vertices, thermodynamic consistency is maintained by
including the thermal drift associated with configuration-dependent
mobility~\cite{lau07}.

To maintain numerical stability, we use adaptive timestepping and
several procedures that help preserve NE mesh quality.
The timestep size is constrained by the maximum deterministic
contribution to the velocity of NE vertices. It is chosen so that
deterministic vertex displacements are no larger than 10\% of the mean
edge length or of the WCA cutoff distance. Mesh quality is maintained
through edge-length tethering, which prevents edges from exceeding
prescribed length bounds, and a local surface tension contribution
that penalizes deviations in triangle area. Edge flips, with
acceptance probabilities determined by changes in discrete tethering
and surface area energies, further promote uniform edge lengths and
triangle areas.

\subsection*{Units and dimensionless parameters}

We use the bending modulus $B$, nuclear radius $R$, and nucleoplasm
viscosity $\mu$ as units of energy, length, and viscosity, so time is
measured in units of $R^3\mu/B$. Table \ref{tab:model_params} lists
the parameter values. Three dimensionless groups control the behavior
of the model,
\begin{gather}
\label{eq:dimensionless_groups}
\lambda=\frac{F_{\text{max}}R}{B},\quad
\rho=\frac{r_{\text{complex}}}{R},\quad
\psi=\frac{R^2\mu v_{\text{poly}}}{B}.
\end{gather}
The force ratio $\lambda$ compares the maximum spindle force to NE
bending force. The size ratio $\rho$ is the SPB-chromatin complex
radius relative to the nuclear radius. The rate ratio $\psi$ compares
the microtubule growth rate to the viscous relaxation rate of the NE.


\begin{table*}
  \footnotesize
    \centering
    \begin{tabular}{p{2in}p{1in}p{3.75in}}
        \textbf{Quantity} & \textbf{Value}  & \textbf{Notes} \\
        \hline
        \textbf{Model units} & &  \\
        Energy $B$ & $4\times 10^{-19}\ \mathrm{J}$ & Estimated \pombe\, NE bending modulus~\cite{lim07}\\
        Length $R$ & $1\ \mathrm{\mu m}$ & Approximate \pombe\, NE radius\\
        Viscosity $\mu$ &  $120\ \mathrm{Pa\!\cdot\!s}$& Estimates in other organisms range from $10$ to $10^3\ \mathrm{Pa\!\cdot\!s}$~\cite{zidovska20a} \\
        \hline
        \textbf{Global simulation parameters} & &  \\
        Maximum timestep $\Delta t_{\text{max}}$ & $ 10^{-5}\ R^3\!\cdot\! \mu/B $ & Upper bound on adaptive timestep, chosen for numerical stability \\
        Thermal energy $k_BT$ & $1.0291\times 10^{-2}\ B$ &  Equivalent to $25^{\circ}\ $C \\
        WCA potential strength $\varepsilon$ &  $12.5\ B$&  Sufficient to prevent steric overlap \\ 
        WCA interaction scale $\sigma$ &  $3.25\times 10^{-2}\ R$& Equivalent to WCA cutoff distance $2^{1/6}\sigma\approx   0.0365\ R$    \\
        \hline
        \textbf{Nuclear envelope} & &  \\
        Surface tension constant $\gamma_0$& $ 32.5\ B/R^2$ & Estimated \pombe\, NE tension~\cite{lim07, lim09} \\
        Area expansion modulus $k_{\text{area}}$& $81.25\ B/R^2$ & Sets magnitude of local surface tension variations \\
        Volume expansion modulus $k_{\text{vol}}$& $2.5\times 10^{4}\ B/R^3$ & Results in less than $1\%$ variation in enclosed volume \\
        Surface drag coefficient $\zeta$&  $1.5\ \mu/R$&  Derived from Stokes' drag law (see supplementary material) \\
        Flip interval $\tau_{\text{flip}}$ & $10^{-5}\ R^3\!\cdot\! \mu/B$  & Controls the frequency of mesh connectivity updates \\
        Flip probability $P_{\text{flip}}$ & $0.3$ & Controls fraction of edges selected as flip candidates per interval \\
        Tether modulus $k_{\text{tether}}$ & $0.1605\ B$ &  Chosen for numerical stability \\
        Tether repulsive singularity $\chi_{\infty}^{\text{rep}}$  & $0.4$ &   Sets minimum edge length, chosen for numerical stability \\
        Tether repulsive onset $\chi_{\text{onset}}^{\text{rep}}$  & $0.8$ &   Chosen for numerical stability \\
        Tether attractive onset $\chi_{\text{onset}}^{\text{att}}$  & $1.2$ &   Chosen for numerical stability \\
        Tether attractive singularity $\chi_{\infty}^{\text{att}}$  & $1.6$ &   Sets maximum edge length, chosen for numerical stability \\
        \hline
        \textbf{Microtubule bundle} & &  \\
        Radius $a$ & $9.5\times 10^{-2}\ R$ &  Approximate fission yeast SPB radius~\cite{ding97}\\
        Polymerization velocity $v_{\text{poly}}$ & $2.5\ B\! \cdot\! R^{-2}\!\cdot\! \mu^{-1}$ &  1/2 spindle elongation rate reported in~\cite{real-calderon26}\\
        Max spindle force $F_{\text{max}}$ & $50$ - $400\ B/R$  & Maximum total extensile force generated by the spindle \\
        Motor force density $f_{\text{motor}}$ & $ 150\ B/R^2$ & Extensile force per unit length of antiparallel microtubule overlap; close to the 50 pN/$\mu$m inferred from protrusions \cite{zhou26a} \\
        \hline
        \textbf{SPB-chromatin complexes} & &  \\
        Radius $r_{\text{complex}}$ & $0.1$ - $0.75\ R$ & Range constrained by SPB size and available nuclear volume \\
        \hline
    \end{tabular}
    \caption{Model parameters.}
    \label{tab:model_params}
\end{table*}

\subsection*{Experimental methods}

\pombe\ strain MB1411 was used for Figures
\ref{fig:mitosis_sim_timelapse}, \ref{fig:spindle_positioning}, and
\ref{fig:protrusions}. Its genotype is \textit{cut11-GFP:ura4+,
  z:adh15:mCherry-atb2:natMX6, leu1-32, his3-D1, ura4-D18, ade6?, h+}.
The cut11-GFP tag labels the NE throughout the cell cycle and the SPBs
from prophase through mid-anaphase.  The fluorescent label for
microtubules was obtained by expressing an
mCherry-$\alpha$-tubulin-chimera at a low level ($\sim$10\% wild type
$\alpha$-tubulin), as described previously~\cite{blackwell17,
  gergely16, edelmaier20, gergely23, gergely23b, jones24,
  zhou26a}. The low-level tubulin labeling helps reduce tag-related
perturbations to microtubule dynamics. Cells were cultured using
standard techniques~\cite{Moreno91}. The strain was constructed using
a genetic cross and random spore analysis to isolate the genotype of
interest. The microscopy images were obtained using live cell
preparation. Cells were grown on YES plates and imaged in EMM liquid
media to reduce background fluorescence and imaged at $25^\circ$
C. The temperature was maintained within $\pm 0.1^\circ$ C using a
CherryTemp temperature controller (Cherry Biotech, Rennes,
France). Spinning-disk confocal microscopy was performed on a Nikon
Eclipse Ti microscope described previously~\cite{blackwell17,
  gergely16, edelmaier20, gergely23, gergely23b, jones24}. 3D
time-lapse images were obtained using the EMGain and laser settings on
the Nikon illumination system and the number of Z-planes indicated
previously~\cite{edelmaier20}.

\section*{Results}

\subsection*{Model validation}

\subsubsection*{Convergence of envelope bending traction and curvature}

\begin{figure}
    \centering
\includegraphics[width=3.25in]{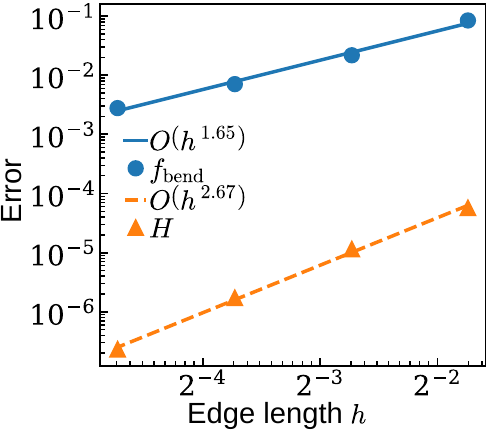}
    \caption{\small 
    Numerical convergence of envelope bending traction and mean curvature. Dimensionless vertex-averaged errors in mean curvature $H$ (triangles) and bending traction $f_{\text{bend}}$ (circles), defined in equations
    \ref{eq:mean_curvature_error} and \ref{eq:bending_stress_error}, are shown as functions of average edge length $h$ for spherical triangulations with
    $|\mathcal{F}|=320$, $1280$, $5120$, and $20480$ faces. Solid and dashed lines show approximate convergence rates.
    }
    \label{fig:curvature_and_bending_convergence}
\end{figure}

To validate the surface curvature and bending traction
discretizations, we compared numerical approximations with known
analytical solutions on the sphere~\cite{meyer03, hildebrandt06}. Using triangulations with
$|\mathcal{F}|=320$, $1280$, $5120$, and $20480$ faces, we computed
the mean curvature and the normal component of the bending traction
\begin{gather}
f_{\text{bend}}
=
-
2B
\left[
\lbo H
+
2H
\left(
H^2-K
\right)
\right]
\end{gather}
at each vertex. We measured the difference between numerical
$(H^{\text{n}}, f^{\text{n}}_{\text{bend}})$ and analytical
$(H^{\text{a}}, f^{\text{a}}_{\text{bend}})=(-1/R, 0)$ solutions using
the vertex-wise average errors~\cite{guckenberger16, guckenberger17}
\begin{gather}
\label{eq:mean_curvature_error}
\varepsilon_{H}
=
\frac{R}{|\mathcal{V}|}
\sum_{i\in\mathcal{V}}
\left|
H^{\text{n}}(\x_i)-H^{\text{a}}(\x_i)
\right|
\\
\label{eq:bending_stress_error}
\varepsilon_{\text{bend}}
=
\frac{R^3}{B |\mathcal{V}|}
\sum_{i\in\mathcal{V}}
\left|
f_{\text{bend}}^{\text{n}}(\x_i)-f_{\text{bend}}^{\text{a}}(\x_i)
\right|
\end{gather}
where $R$ is the sphere's radius. The error decreases with increasing
mesh resolution, which we quantify using the average edge length $h$,
as expected. We find approximate $O(h^{2.67})$ and $O(h^{1.65})$
convergence in $H$ and $f_{\text{bend}}$, respectively
(Fig.~\ref{fig:curvature_and_bending_convergence}).

\subsubsection*{Analysis of envelope thermal fluctuation spectra}

\begin{figure}
    \centering
\includegraphics[width=3.25in]{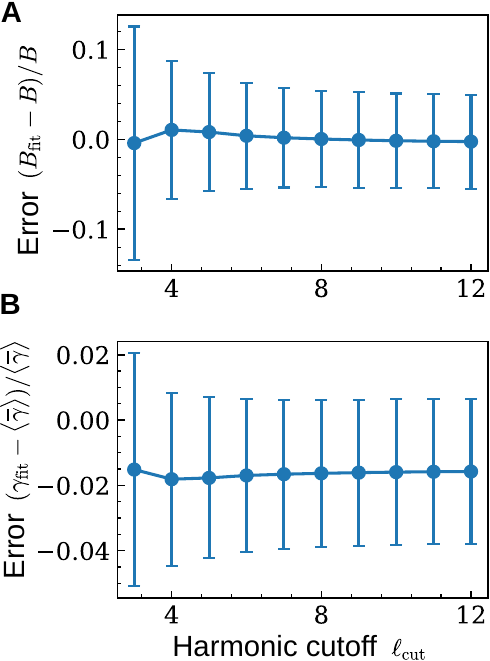}
    \caption{\small 
    Validation of envelope thermal fluctuation spectra. (A) Relative error in the fitted bending modulus $B_{\text{fit}}$. (B) Relative error in the fitted uniform tension $\gamma_{\text{fit}}$. Errors are shown as functions of the maximum spherical harmonic degree $\ell_{\text{cut}}$ included in the fit and were computed relative to the prescribed bending modulus $B$ and the surface- and time-averaged tension $\texpect{\saverage{\gamma}}$. The measured fluctuation spectra were fitted to Eq. \ref{eq:predicted_spectrum} using unweighted nonlinear least squares. Ensemble averages were approximated using 200 configurations sampled at intervals of $\Delta t_{\text{sample}}=0.06$. Data are represented as the mean$\pm$standard error across 12 simulations.
    }
    \label{fig:stiffness_and_tension_fit}
\end{figure}

Analysis of thermal fluctuations provides further numerical validation
of the NE model. The theory of Milner and Safran~\cite{safran83,
  milner87} relates the equilibrium shape fluctuations of a nearly
spherical membrane to its bending stiffness and a spatially uniform
measure of surface tension. We estimated the stiffness and tension of
simulated nuclei from their fluctuation spectra and compared the
estimates to known input parameters of the model. This analysis is
related to flicker spectroscopy-based techniques for quantifying
mechanical properties of membranes~\cite{faucon89, rautu17} and
similar methods have been used to validate other dynamically
triangulated membrane models~\cite{penic15}.

Near equilibrium, the envelope can be described in terms of a small
perturbation of a reference sphere
\begin{gather}
\r(\theta,\phi, t)
=
\mathbf{c}(t)
+
R_0\left[
1+u(\theta,\phi, t)
\right]\e_r(\theta, \phi),
\end{gather}
where $\mathbf{c}(t)$ is the instantaneous NE centroid, the
reference spherical radius $R_0$ is determined by the envelope's
initial enclosed volume $V_0=4\pi R_0^3/3$, $\e_r(\theta, \phi)$ is
the radial unit vector, and $u$ is the dimensionless radial
perturbation. We expand $u$ in terms of spherical harmonics
\begin{gather}
u(\theta, \phi, t)
=
\sum_{\ell=0}^\infty\sum_{m=-\ell}^\ell
u_{\ell m}(t)Y_{\ell m}(\theta,\phi),
\end{gather}
and define the $m$-averaged mean-squared fluctuation spectrum
\begin{gather}
\label{eq:mean_sqr_spectrum}
C_{\ell}
=
\frac{1}{2\ell+1}\sum_{m=-\ell}^{\ell}
\texpect{|u_{\ell m}|^2}.
\end{gather}
The Milner-Safran analysis predicts~\cite{safran83, milner87}
\begin{gather}
\label{eq:predicted_spectrum}
C_{\ell}(B, \Sigma)
=
\frac{k_BT}{B(\ell-1)(\ell+2)(\Sigma + \ell(\ell+1))}
\end{gather}
where $\Sigma=\gamma_{\text{MS}} R_0^2/B$ is related to a uniform
tension $\gamma_{\text{MS}}$ and angle brackets denote an ensemble
average.

For each simulation, we estimated the spectrum by averaging over $N_t$
sampled configurations,
\begin{gather}
\label{eq:observed_spectrum}
\widehat{C}_\ell
=
\frac{1}{N_t(2\ell + 1)}
\sum_{k=1}^{N_t}
\sum_{m=-\ell}^\ell
|u_{\ell m}(t_k)|^2,
\end{gather}
with coefficients $u_{\ell m}(t_k)$ approximated by numerical
integration over the discrete surface. A nonlinear least squares fit
of $\widehat{C}_\ell$ to the theoretical spectrum in Eq.
\ref{eq:predicted_spectrum} over degrees $2\leq \ell\leq
\ell_{\text{cut}}$ yields estimates $B_{\text{fit}}$ and
$\Sigma_{\text{fit}}$ with corresponding uniform tension
$\gamma_{\text{fit}}=B_{\text{fit}}\Sigma_{\text{fit}}/R_0^2$. In the
Milner and Safran formulation, $\gamma_{\text{MS}}$ is a constant Lagrange
multiplier chosen to fix the ensemble averaged membrane area. To
compare the theoretical prediction with our simulations, we use the
surface- and time-averaged tension, $\langle\bar\gamma\rangle$, as a
reference for the fitted value $\gamma_{\text{fit}}$.

Across 12 runs, the mean fitted bending stiffness approached the
prescribed value as $\ell_{\text{cut}}$ increased, with less than
$1\%$ discrepancy at the larger cutoffs tested
(Fig.~\ref{fig:stiffness_and_tension_fit}). The mean fitted
tension remained approximately $2\%$ below
$\texpect{\saverage{\gamma}}$.

\subsection*{Model of mitotic NE-spindle interactions}

Mitosis in \pombe\ can be divided into three phases. During phase 1,
the bipolar spindle forms and grows to span the diameter of the
nucleus. In phase 2, spindle growth slows and sister chromatids
segregate to opposite spindle poles. During phase 3 (anaphase B), the
spindle rapidly elongates before reaching its maximum length and
disassembling~\cite{tolic-norrelykke04, fu09, real-calderon26}. Here
we model phases 1 and 3, where spindle and NE mechanics are
most directly coupled. Phase 2 chromosome segregation is not modeled
directly, but is accounted for by varying the size of SPB-chromatin
complexes between our phase 1 and phase 3 simulations.

\subsubsection*{Dynamics of initially displaced spindles}

\begin{figure*}
    \centering
\includegraphics[width=6.75in]{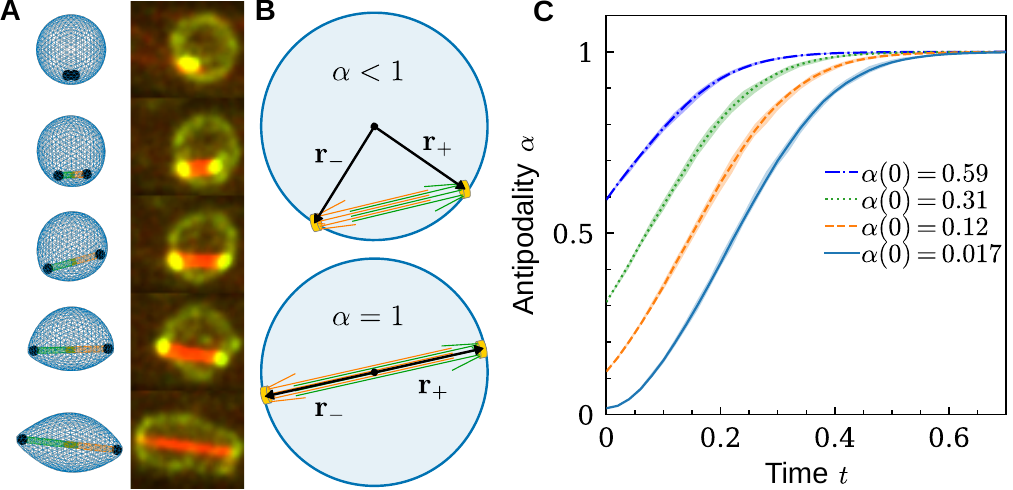}
    \caption{\small 
    Growth-driven spindle centering and alignment.
    (A) Simulated (left) and experimental (right)
    time-lapse sequences showing centering and alignment of 
    initially mispositioned spindles.
    Experimental images show microtubules labeled with mCherry-Atb2
    (orange) and the NE and SPBs labeled with Cut11-GFP
    (green). (B) SPB antipodality $\alpha$ is computed from the 
    positions $\r_{\pm}$ of the SPBs relative to the NE centroid. 
    (C) Evolution of SPB antipodality $\alpha(t)$ for
    four ensembles with different starting values $\alpha(0)$ in an initially
    spherical nucleus. Curves show ensemble means and shaded
    regions show spread across simulations. 
    }
    \label{fig:spindle_positioning}
\end{figure*}

During its initial formation, the spindle is often
displaced from the nuclear center
(Fig.~\ref{fig:spindle_positioning}). However, proper spindle assembly
requires that the SPBs move toward antipodal points on the NE. We
performed simulations with initially mispositioned spindles and
quantified relaxation toward centered configurations with antipodal
SPBs. Because spindle positioning and alignment occur prior to
chromosome segregation, we choose $\rho=0.15$ for these simulations.
This choice corresponds to an SPB-chromatin complex size only slightly larger than the SPB. To
account for the fact that SPBs remain embedded in the NE during closed
mitosis, simulations were initialized with a surface-to-surface
separation between SPBs and NE small enough that the steric potential
(Eq.~\ref{eq:wca_potential}) is non-zero.

We quantify SPB antipodality by
\begin{equation}
    \alpha
    =
    \frac{1}{2}
    \left(
    1
    -
    \frac{
    \r_{+}\cdot \r_{-}
    }{
    |\r_{+}| |\r_{-}|
    }
    \right)
\end{equation}
where $\r_{\pm} = \X_{\text{spb}, \pm}-\mathbf{c}$ is the position of
the SPB-chromatin complex associated with the spindle endpoint
$\X_{\text{mt}}\pm L\vecU_z/2$, relative to the NE centroid
$\mathbf{c}$ (Fig.~\ref{fig:spindle_positioning}B). The quantity
$\alpha$ ranges between 0 and 1, with $\alpha=0$ when SPBs lie in the same direction from $\mathbf{c}$ and $\alpha=1$ when
they lie in opposite directions. For a range of initial
conditions, we observed that spindle elongation drives initially
mispositioned spindles toward configurations in which the SPBs are antipodal
(Fig.~\ref{fig:spindle_positioning}C) and the spindle midpoint
coincides with the NE centroid. The spindle axis and principal axis of
the NE also approach alignment.

\subsubsection*{Anaphase nuclear shape}

\begin{figure*}
    \centering
\includegraphics[width=6.75in]{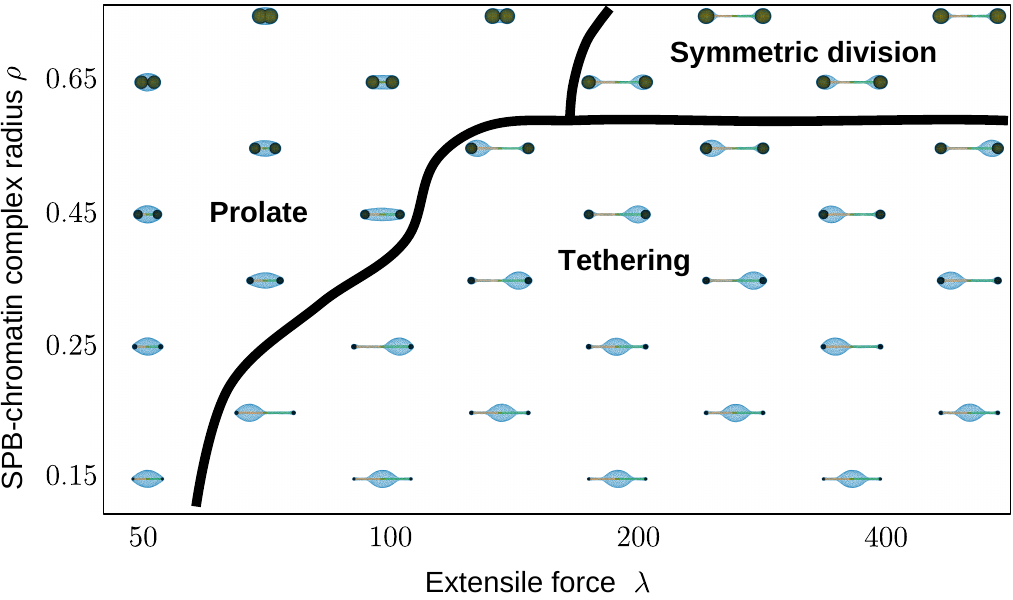}
\caption{\small 
Phase diagram of anaphase nuclear shapes. Characteristic nuclear 
shape attained following spindle elongation is shown as
a function of dimensionless extensile force
$\lambda=F_{\text{max}}R/B$ and dimensionless SPB-chromatin complex radius
$\rho=r_{\text{complex}}/R$ for systems with $L_{\text{half}}=3R$. 
Black curves indicate approximate boundaries between morphological regimes.
The formation of dumbbell shapes compatible with symmetric nuclear
division is only observed for sufficiently large values of $\lambda$
and $\rho$. Low extensile force produces undivided prolate nuclei, while small SPB-chromatin
complex size leads to thin NE tethers. 
Near the boundary of the
tethering regime, nuclei exhibit strongly asymmetric
configurations.
  }
    \label{fig:phase_daigram}
\end{figure*}

To simulate NE shape dynamics during anaphase B, we first noted that
during anaphase the spindle midzone has approximately constant length
\cite{ward14}. Equation~\ref{eq:mt_overlap_length_evolution} therefore
implies $v_{\text{poly}}\approx \frac{1}{2}\dv[L]{t}$. In our anaphase
B simulations, we choose the polymerization velocity $v_{\text{poly}}$
of each half-bundle to be approximately half the reported spindle
elongation rate~\cite{real-calderon26}. Simulations of the elongating
spindle reproduced features of both successful and aberrant closed
mitosis in \pombe, including the sequence of shapes from spherical to
dumbbell (Fig.~\ref{fig:mitosis_sim_timelapse}).  When the
SPB-chromatin complex size is small, our simulations formed NE tethers
(Fig.~\ref{fig:protrusions}).


To map the landscape of possible nuclear
morphologies, we fix a maximum microtubule length,
$L_{\text{half}}=3R$, at which to stop microtubule polymerization and examine quasi-equilibrium configurations for systems with a range of $\rho=r_{\text{complex}}/R$ and $\lambda=F_{\text{max}}R/B$ values.
The resulting phase diagram (Fig.~\ref{fig:phase_daigram}) showed
three regimes. For small $\lambda$, spindle-generated forces produce
elongated prolate nuclei. Increasing $\lambda$ led to symmetric shapes
only when spindle forces are distributed over a sufficiently large
region of the NE. For small $\rho$, spindle elongation instead
produced nuclei with either a single narrow NE tether or a symmetric pair of tethers. Near the boundaries of the tethering regime, we observed exclusively asymmetric, single-tether configurations.

\subsubsection*{Tethers and protrusions}
\begin{figure*}
    \centering
\includegraphics[width=6.75in]{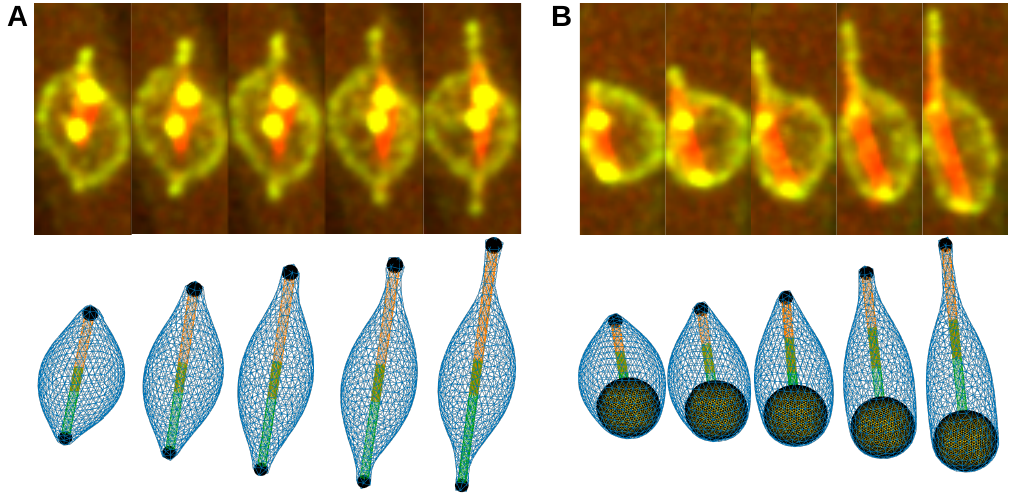}
    \caption{\small 
    Formation of symmetric and asymmetric nuclear protrusions.
    (A) Symmetric protrusions. Top: experimental time-lapse images
    with microtubules labeled with mCherry-Atb2 (orange) and the NE
    and SPBs labeled with Cut11-GFP (green). Microtubules push out the
    NE at both ends of the spindle. Bottom: simulation with small
    SPB-chromatin complexes at both spindle ends, which forms a narrow
    NE tether at each pole. (B) Asymmetric protrusion. Top:
    experimental time-lapse images in which a single protrusion forms
    at one end of the spindle. Bottom: simulation with SPB-chromatin
    complexes of unequal size, in which a single tether forms at the
    smaller complex.
    }
  \label{fig:protrusions}
\end{figure*}
 
The morphologies observed with symmetric tethering are similar to
previous observations of nuclear deformation by small microtubule
bundles or the spindle if chromosome segregation is disrupted
\cite{lim07, gergely16}. We also observed symmetric and asymmetric
tethering experimentally in a genetic background in which spindle
microtubule minus ends can detach from the SPBs and form protrusions
that deform the NE (Fig. \ref{fig:protrusions}) \cite{toya07,
  yukawa15, syrovatkina15, gergely23, jones24, zhou26a}. In this case,
the NE is deformed by a single or small bundle of microtubules rather
than the SPB and chromatin. The protrusions that form are consistent
with predictions of the model for small $\rho$.

\subsubsection*{Nuclear area and volume variations during mitosis}

\begin{figure}
    \centering
\includegraphics[width=3.25in]{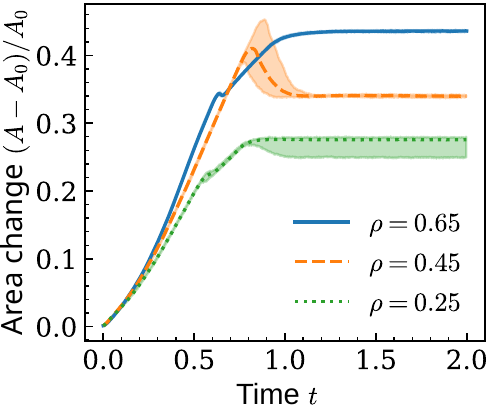}
    \caption{\small 
    Nuclear envelope area increase during mitosis. Relative changes in NE surface area $(A-A_0)/A_0$ are shown for ensembles with dimensionless SPB-chromatin
    complex radii $\rho=0.25, 0.45, 0.65$, where $A_0$ is the initial NE area. Curves show ensemble
    means and shaded regions show ensemble spread across simulations.
    }
    \label{fig:area_change_during_mitosis}
\end{figure}

During mitosis in \pombe, enclosed nuclear volume remains
approximately constant while nuclear surface area
increases~\cite{yam11, lim07}. Our simulations gave similar results,
with $<1\%$ variation in enclosed volume and surface area increases of
$25\%$ to $44\%$ (Fig.~\ref{fig:area_change_during_mitosis}). Larger
increases in surface area occurred for symmetric nuclear shapes, while
asymmetric shapes generally required lower area increases. This
distinction was noticeable for $\rho=0.45$, which led to asymmetric
nuclei, and $\rho=0.65$, which led to symmetric shapes. We note that
these values include the area of the entire nucleus, including the
nuclear bridge connecting the two sides.

The larger variation in final surface area for systems with
$\rho=0.25$ (Fig.~\ref{fig:area_change_during_mitosis}) reflects the
range of shapes that occurred at small radii, with about $7\%$ of
simulations showing single tethers and $93\%$ forming two tethers. In
contrast, simulations with $\rho=0.45$ and $\rho=0.65$ all reached
shapes with similar surface areas for each value of $\rho$.

\subsubsection*{Reduced-order description of nuclear shape
  trajectories}

\begin{figure*}
    \centering
\includegraphics[width=6.75in]{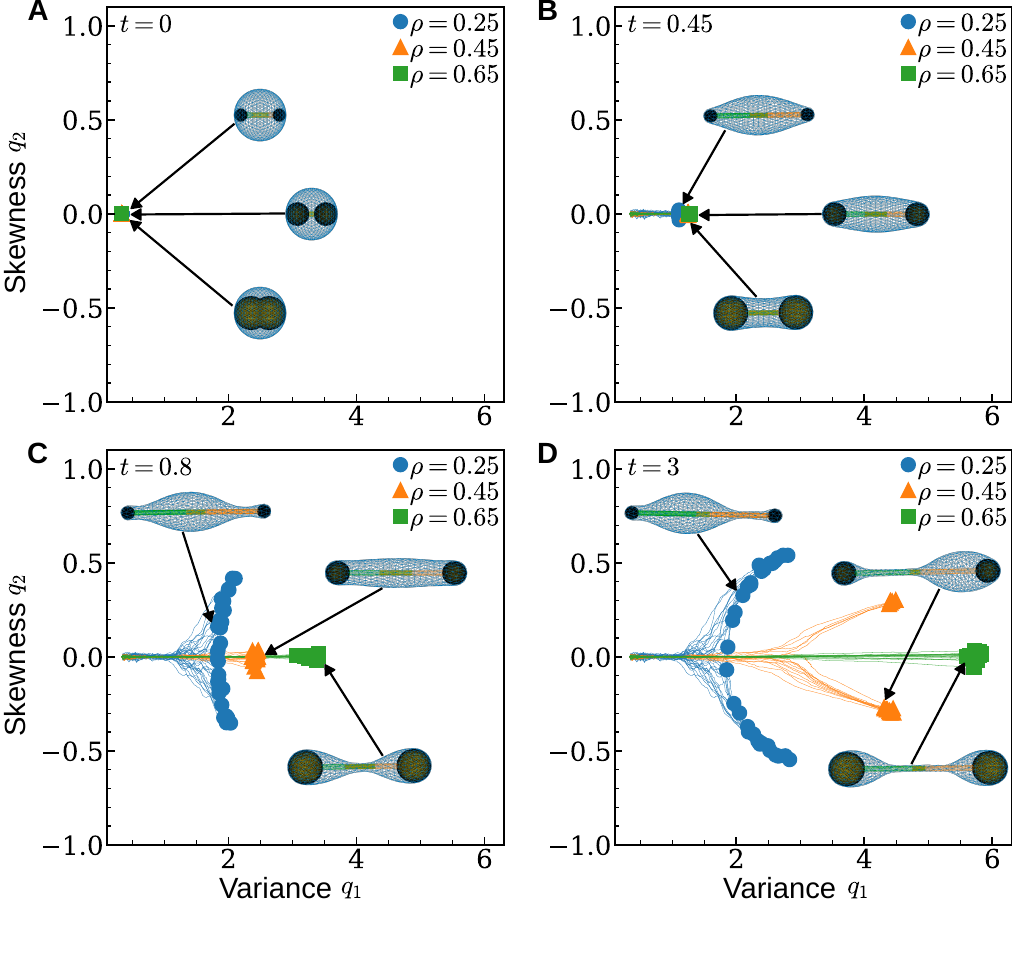}
    \caption{\small 
    Spontaneous symmetry breaking during spindle elongation. Ensembles of nuclear shape trajectories $(q_1(t), q_2(t))$ with initially symmetric configurations and dimensionless SPB-chromatin complex size $\rho=0.25$, $0.45$, $0.65$ reveal the existence of an attractor corresponding to symmetric division which is stable for sufficiently large $\rho$. Panels (A-D) show trajectories up to simulation times $t=0$, $0.45$, $0.8$, and $3$, respectively. Markers indicate instantaneous values of $(q_1(t), q_2(t))$ and insets show representative nuclear configurations. 
    Trajectories initially evolve along a common path before diverging
    toward different long-time configurations. As $\rho$ approaches
    the critical value for symmetric division from above, new
    attractors emerge and the system exhibits spontaneous symmetry
    breaking as trajectories with $\rho$ below the critical value
    evolve toward asymmetric nuclear configurations.
    }
    \label{fig:bifurcations}
\end{figure*}

To characterize morphological transitions in the envelope-spindle system, we introduce two reduced-order shape observables
\begin{gather}
    q_1 = \saverage{\delta z^2}
    ,\quad
    q_2 =\frac{\saverage{\delta z^3}}{\left(\saverage{\delta z^2}\right)^{3/2}}
\end{gather}
where
\begin{gather}
    \delta z\left(\x\right)=\vecU_z\cdot\left(\x - \X_{\text{mt}}\right)/R,
\end{gather}
and $\saverage{\,\cdot\,}$ denotes an average over the envelope
surface
\begin{gather}
    \saverage{ f\left(\x\right)} = \frac{\int_M f\left(\x\right)\dd{A}}{\int_M\dd{A}}.
\end{gather}
The quantities $q_1$ and $q_2$ characterize the second and third
moments of the NE surface area distribution along the spindle axis,
measured relative to the spindle midpoint. Larger values of $q_1$ indicate more elongated nuclei. Shapes with reflection symmetry about the spindle midplane have $q_2=0$, while nonzero $q_2$ indicates shape asymmetry. 

The observables $q_1$ and $q_2$ reveal qualitatively distinct dynamics
during spindle elongation for nuclei with different values of $\rho$
(Fig.~\ref{fig:bifurcations}). As half-bundles grow to a maximum
length $L_{\text{half}}=3R$ in initially symmetric systems,
trajectories with $\rho=0.65$ remain
approximately symmetric with $q_2$ near zero, while trajectories with
$\rho=0.45$ and $\rho=0.25$ branch toward nonzero $q_2$.  Initially
symmetric shapes therefore become reflection-asymmetric even though no
difference is imposed between the two SPB-chromatin complexes. While
not explored in detail here, similar symmetry-breaking bifurcations
were also observed when varying the dimensionless force $\lambda$
(Fig.~\ref{fig:phase_daigram}) or the degree of asymmetry between the
two SPB-chromatin complexes (Fig.~\ref{fig:protrusions}).

\section*{Discussion}

\subsection*{Envelope-spindle interactions promote spindle centering and alignment}

As the bipolar spindle begins to elongate and SPBs separate in early
mitosis, mechanical interactions with the NE serve as an
error-correcting mechanism that centers the spindle within the
nucleus. In an initially spherical nucleus, spindle elongation
produces small deformations of the NE that establish a principal axis
for the envelope approximately aligned with the spindle. As elongation
continues, NE and SPB-chromatin complex interactions generate forces
and torques that maintain NE-spindle alignment and reduce the
separation between the spindle midpoint and envelope
centroid. Approximate axisymmetry of the coupled envelope-spindle
system therefore emerges dynamically, with spindle elongation
deforming the NE while the NE influences spindle position and
orientation.  This centering and alignment is emergent, without added
forces or activity designed to center or align the spindle.  This
suggests that mechanical interactions with the NE can contribute to
proper spindle positioning during early mitosis, prior to chromosome
segregation.  Our experimental observations are consistent with this
mechanism.

\subsection*{Spindle force magnitude and distribution govern nuclear morphology}

In our model, successful closed mitosis depends not only on the
magnitude of spindle-generated forces, but also on their distribution
over portions of the NE substantially larger than the spindle pole
bodies, as found in previous modeling work \cite{castagnetti15, zhu16,
  fang20}.  We found two characteristic failures of NE shape change:
insufficient spindle force produces elongated nuclei without forming a
dumbbell shape, while larger forces distributed over small regions of
the NE produce narrow envelope tethers
(Fig.~\ref{fig:phase_daigram}). These tethers resemble the thin NE
protrusions observed experimentally \cite{lim07, zhou26a}. Increasing
the size of the SPB-chromatin complexes distributes forces over larger
regions of the envelope, driving the spherical-to-dumbbell shape transition if the spindle force is sufficiently large.

The variety of shapes observed in the tethering regime shows that
relatively small changes in the magnitude or spatial distribution of
spindle forces can cause qualitatively distinct changes in nuclear
shape. We noted that strongly asymmetric, single-tether shapes were
only observed near the boundary with symmetric division. We also found that the total extensile force, rather than the motor force density itself, is the more relevant quantity governing nuclear morphology. 

Our findings are in qualitative agreement with results of previous
axisymmetric models~\cite{lim07, castagnetti15, zhu16,
  fang20}. Although the SPBs are relatively small ($\sim$200 nm),
chromatin clustered near the SPBs can significantly increase the area
of NE over which spindle-generated forces are distributed. This
provides a mechanism by which chromatin organization can influence
nuclear division independently of spindle-generated force magnitude
and suggests that changes in chromatin organization could impair
symmetric nuclear division even when the spindle generates sufficient
extensile force.

\subsection*{Nuclear symmetry, stability, and dynamics}

This work extends previous NE modeling in fission yeast
to non-axisymmetric nuclear shapes and dynamics. Axisymmetric membrane
models have provided important insights into the mechanics of NE shape
\cite{lim07, lim09, castagnetti15, zhu16}, with the advantage of model
simplicity but the disadvantage of restricting the shapes and dynamics
that can be modeled. For the coupled NE-spindle system, axisymmetry
requires a common symmetry axis for the spindle and nucleus. This
excludes shapes with laterally displaced or misaligned spindles
(Fig.~\ref{fig:spindle_positioning}). Here, the model allows the
spindle to translate and rotate within the nucleus, allowing modeling
of additional aspects of nuclear dynamics.

Imposing axisymmetry can also affect predictions of nuclear shape
stability and transitions between shapes. Previous work has identified
sequences of equilibrium axisymmetric shapes~\cite{lim07, lim09,
  castagnetti15, zhu16} and dynamical trajectories~\cite{fang20} which
include transitions from spherical to dumbbell or tethered
configurations.  A shape that minimizes energy in an axisymmetric
model can be unstable to perturbations outside the axisymmetric
subspace, potentially altering dynamics from the sequence of shapes
predicted by the constrained model. Related membrane models exhibit
bifurcations from axisymmetric to nonaxisymmetric
shapes~\cite{heinrich93, meiners26}, and transitions between
axisymmetric membrane configurations need not imply axisymmetry of
intermediate shapes~\cite{bahrami17}. Access to non-axisymmetric
deformations can also shift mechanical thresholds and critical
values~\cite{vasan20}.

Nuclei have non-axisymmetric thermal fluctuations, so dynamical modes
excluded by imposed axisymmetry are continually excited in physical
nuclei. Our simulations allow non-axisymmetric deformations and
thermal fluctuations, yet still recover the qualitative dependence of
nuclear morphology on the spatial distribution of spindle forces
predicted by earlier axisymmetric models. In particular, sufficiently
large spindle forces distributed over sufficiently large
chromatin-envelope contact regions produce transitions from spherical
to symmetric dumbbell configurations, while more localized force
transmission leads to the formation of narrow NE tethers. Thus, the
principal mechanical regimes identified by previous models
\cite{lim07, lim09, castagnetti15, zhu16, fang20} persist in our
non-axisymmetric model.  Whether non-axisymmetric perturbations alter
the stability of these configurations or the parameter values
separating shape regimes would be of interest to study in future work.
Determining the influence of axisymmetry constraints on the stability
and transitions of nuclear configurations will require comparison of
constrained and unconstrained dynamics within the same mechanical
model.

\subsection*{Spontaneous symmetry breaking in closed mitosis}

Our quantification with reduced-order variables shows that the
stability of symmetric nuclear shapes depends on the size of the
SPB-chromatin complexes. For sufficiently large radii, nuclear shapes
remain approximately symmetric with the asymmetry measure $q_2$
remaining small in magnitude during spindle elongation, despite
potentially destabilizing thermal fluctuations. When the complex
radius decreases to $\rho\approx 0.5$, symmetric shapes lose
stability.  Initially spherical nuclei evolve toward asymmetric shapes
with either positive or negative $q_2$. Because the two SPB-chromatin
complexes are identical, neither asymmetric configuration is
favored. Instead, thermal fluctuations perturb the system away from a
symmetric state and toward one of the two asymmetric configurations
with equal probability.

The differences in NE surface area between symmetric and asymmetric
shapes suggest that surface tension may also play a role in
determining the onset of symmetry breaking. At fixed enclosed volume,
symmetric shapes with two equally sized lobes typically require more
surface area than shapes with unequal lobes
(Fig.~\ref{fig:area_change_during_mitosis}). Thus we predict higher
surface tension will favor asymmetric shapes and shift the critical
SPB-chromatin complex radius for symmetry breaking toward larger
values. A similar effect is expected in the tethering regime, where
single-tether shapes have smaller surface areas than those with two
tethers. Increasing surface tension should therefore favor asymmetric
nuclear shapes, as long as surface tension does not prevent spindle
elongation. Determining how surface tension affects the stability of
symmetric shapes and the critical parameter values separating these
regimes would be of interest in future work. 

\section*{Author Contributions}

W.L., E.C., and M.D.B. designed the research and developed the
model. W.L. wrote the simulation code, performed the simulations,
performed model validation, and analyzed the simulation results.
Z.R.G. performed experimental measurements and analyzed the
experimental data. W.L.  wrote the first draft of the manuscript. All
authors edited the manuscript and approved the final version.

\section*{Acknowledgments}

We thank Andrew Berger, Mary Williard Elting, Dick McIntosh, Daniel
Steckhahn, and Bojun Zhou for helpful discussions. This work was
supported by NSF grant MCB 2133243 to M.D.B.  We acknowledge the Light
Microscopy Core Facility in Porter Biosciences at the University of
Colorado Boulder (RRID:SCR\_018993) for help and advice with
microscopy.

\section*{Declaration of Interests}

The authors declare no competing interests.
 
\bibliography{references}


\section*{Supplementary Material}

An online supplement to this article can be found by visiting BJ
Online at \url{http://www.biophysj.org}.

\end{document}


\begin{frontmatter}
\end{frontmatter}

\section*{Notation}

Table \ref{tab:notation} lists the symbols used in the main text and in this section.

{\small
\begin{longtable}{p{1.3in}p{3.3in}p{1.7in}}
\caption{Notation. A symbol with more than one meaning has one row per
  meaning. ``Main'' refers to equations in the main text.}
\label{tab:notation}\\
\textbf{Symbol} & \textbf{Meaning} & \textbf{Defined in} \\
\hline
\endfirsthead
\textbf{Symbol} & \textbf{Meaning} & \textbf{Defined in} \\
\hline
\endhead
\hline
\endfoot
\multicolumn{3}{l}{\textbf{General notation}} \\
$\mathbf{I}$ & $3\times3$ identity matrix & Envelope mechanics \\
$\E_x$, $\E_y$, $\E_z$ & Fixed orthonormal basis for $\mathbb{R}^3$ & Fig. \ref{fig:mt_bundle_config_and_velocities} \\
$\hat{\mathbf{a}}$ & Antisymmetric matrix satisfying $\hat{\mathbf{a}}\mathbf{b}=\mathbf{a}\times\mathbf{b}$ & Langevin dynamics \\
$\delta_{ij}$ & Kronecker delta & Discretization \\
$\delta(t-t')$ & Dirac delta in time & Eq. \ref{eq:f_thermal_covariance} \\
$k_BT$ & Thermal energy & Eq. \ref{eq:f_thermal_covariance} \\
$\langle\cdot\rangle$ & Ensemble expectation & Envelope mechanics \\
\multicolumn{3}{l}{\textbf{Surface geometry}} \\
$M$ & Smooth nuclear envelope surface & Envelope model \\
$s=(s^1,s^2)$ & Material surface coordinates & Envelope model \\
$\r(s,t)$ & Embedding of $M$ in $\mathbb{R}^3$ & Envelope model \\
$\e_\alpha$, $\e^\alpha$ & Covariant and contravariant tangent basis vectors & Envelope model \\
$\n$ & Outward unit normal & Envelope model \\
$\nabla_\alpha$, $\nabla_M$ & Covariant derivative, surface gradient & Envelope model \\
$\lbo$ & Laplace-Beltrami operator (negative semidefinite) & Envelope model \\
$H$ & Mean curvature (negative on a sphere) & Envelope model \\
$K$ & Gaussian curvature & Envelope model \\
$P$, $\partial P$ & Surface patch and its boundary & Envelope model \\
$\mathbf{u}$ & Outward unit conormal on $\partial P$ & Envelope model \\
$\dd{A}$, $\dd{s}$ & Surface area and boundary length elements & Envelope model \\
\multicolumn{3}{l}{\textbf{Envelope mechanics}} \\
$\stensor$ & Surface stress tensor & Envelope model \\
$\stensor^\alpha$ & Vector-valued components of the surface stress tensor & Envelope model \\
$\stensor_{\text{bend}}$, $\stensor_{\text{tension}}$ & Bending and tension stresses & Envelope model \\
$U_{\text{bend}}$ & Bending energy & Eq. \ref{eq:bending_energy}, Main Eq. \ref{M-eq:envelope_energy} \\
$B$ & Bending modulus & Eq. \ref{eq:bending_energy} \\
$\gamma$ & Surface tension & Envelope model \\
$\gamma_0$ & Uniform part of surface tension & Envelope model \\
$\gamma_{\text{local}}$ & Strain-dependent part of surface tension & Envelope model \\
$U_{\text{vol}}$ & Volume constraint potential & Envelope model \\
$V$, $V_0$ & Enclosed volume and reference volume & Envelope model \\
$k_{\text{vol}}$ & Volume expansion modulus & Envelope model \\
$p$ & Effective pressure & Envelope model \\
$\f_{\text{viscous}}$ & Viscous traction & Eq. \ref{eq:f_viscous} \\
$\zeta$ & Drag coefficient per unit area & Eq. \ref{eq:f_viscous} \\
$\vecv$ & Local surface velocity & Eq. \ref{eq:f_viscous} \\
$\F_{\text{Stokes}}$ & Stokes drag force on a sphere & Surface drag coefficient \\
$\f_{\text{thermal}}$ & Thermal traction & Eq. \ref{eq:f_thermal_mean} \\
$\delta_M(s,s')$ & Dirac delta on $M$ & Eq. \ref{eq:f_thermal_covariance} \\
$\f_{\text{contact}}$ & Contact traction from spindle and chromatin & Eq. \ref{eq:traction_balance} \\
\multicolumn{3}{l}{\textbf{Envelope discretization}} \\
$M^h$ & Triangulated envelope surface & Discretization \\
$h$ & Mesh length scale & Discretization \\
$\mathcal{V}$, $\mathcal{E}$, $\mathcal{F}$ & Sets of vertices, edges, and faces & Discretization \\
$|\mathcal{V}|$, $|\mathcal{F}|$ & Numbers of vertices and faces & Main, model validation \\
$\mathcal{V}_i$, $\mathcal{E}_i$, $\mathcal{F}_i$ & Vertices adjacent to, edges incident on, and faces incident on vertex $i$ & Discretization \\
$\x_i$, $\mathbf{v}_i$ & Position and velocity of vertex $i$ & Discretization \\
$P_i^h$, $P_i$ & Dual cell of vertex $i$ and its corresponding smooth patch & Fig. \ref{fig:triangle_mesh_fig} \\
$a_i^*$ & Dual cell area & Eq. \ref{eq:dual_cell_drag} \\
$\boldsymbol{\Gamma}_i$ & Drag tensor of vertex $i$ & Eq. \ref{eq:dual_cell_drag_tensor} \\
$\dd{A^h}$ & Area element on $M^h$ & Discretization \\
$\n^h$ & Piecewise-constant unit normal on $M^h$ & Discretization \\
$\n_f$ & Unit normal of face $f$ & Eq. \ref{eq:F_tension} \\
$\tilde{\n}_i$ & Unnormalized vertex normal & Discretization \\
$\n_i$ & Unit normal at vertex $i$ & Eq. \ref{eq:vertex_normal} \\
$\cotlap$ & Cotangent Laplacian & Eq. \ref{eq:cotan_formula} \\
$\theta_{ij}^{\pm}$ & Angles opposite edge $(i,j)$ & Eq. \ref{eq:cotan_formula} \\
$H_i$ & Discrete mean curvature at vertex $i$ & Eq. \ref{eq:discrete_mean_curvature} \\
$K_i$ & Discrete Gaussian curvature at vertex $i$ & Eq. \ref{eq:discrete_gaussian_curvature} \\
$\phi_{if}$ & Interior angle of face $f$ at vertex $i$ & Eq. \ref{eq:discrete_gaussian_curvature} \\
$\gamma_f$ & Surface tension of face $f$ & Eq. \ref{eq:face_tension} \\
$a_f$ & Area of face $f$ & Eq. \ref{eq:face_tension} \\
$a_0$ & Reference face area & Eq. \ref{eq:face_tension} \\
$k_{\text{area}}$ & Area expansion modulus & Eq. \ref{eq:face_tension} \\
$\gamma^h$ & Piecewise-constant tension field & Discretization \\
$\mathbf{u}^h$, $\dd{s^h}$ & Unit conormal and length element on $\partial P_i^h$ & Discretization \\
$\mathbf{h}_{if}$ & Directed edge opposite $i$ in face $f$ & Eq. \ref{eq:F_tension} \\
$U_{\text{area}}$ & Discrete area energy & Eq. \ref{eq:discrete_area_energy} \\
$A^h$ & Area of $M^h$ & Eq. \ref{eq:discrete_area_energy} \\
$V^h$, $p^h$ & Enclosed volume and effective pressure of $M^h$ & Discretization \\
$U^h_{\text{vol}}$ & Discrete volume penalty energy & Discretization \\
$\F_{\text{bend},i}$ & Bending force on vertex $i$ & Eq. \ref{eq:F_bend} \\
$\F_{\text{tension},i}$ & Tension force on vertex $i$ & Eq. \ref{eq:F_tension} \\
$\F_{\text{pressure},i}$ & Pressure force on vertex $i$ & Eq. \ref{eq:F_pressure} \\
$\F_P$ & Total thermal force on surface patch $P$ & Discretization \\
$A_{P\cap P'}$ & Area of the intersection of two patches & Discretization \\
$\F_{\text{thermal},i}$ & Thermal force on vertex $i$ & Eq. \ref{eq:F_thermal} \\
$\boldsymbol{\xi}_i$ & Gaussian white noise at vertex $i$ & Eq. \ref{eq:F_thermal} \\
$\F_{\text{drift},i}$ & Thermal drift force on vertex $i$ & Eq. \ref{eq:F_drift} \\
$U_{\text{tether}}$ & Tethering energy of an individual edge & Eq. \ref{eq:tethering_energy} \\
$k_{\text{tether}}$ & Tether modulus & Eq. \ref{eq:tethering_energy} \\
$G$ & Dimensionless tethering energy profile & Discretization \\
$h_0$ & Reference edge length & Eq. \ref{eq:tethering_energy} \\
$\chi_{ij}$ & Dimensionless length of edge $(i, j)$ & Eq. \ref{eq:F_tether} \\
$\chi$ & Dimensionless edge length & Eq. \ref{eq:tethering_energy} \\
$\chi^{\text{rep}}_{\text{onset}}$, $\chi^{\text{att}}_{\text{onset}}$ & Dimensionless edge lengths where tethering begins & Eq. \ref{eq:tethering_energy} \\
$\chi^{\text{rep}}_{\infty}$, $\chi^{\text{att}}_{\infty}$ & Dimensionless edge lengths where tethering energy diverges & Eq. \ref{eq:tethering_energy} \\
$h_{\text{min}}$, $h_{\text{max}}$ & Minimum and maximum allowable edge lengths & Discretization \\
$\F_{\text{tether},i}$ & Total tethering force on vertex $i$ & Eq. \ref{eq:F_tether} \\
$\tau_{\text{flip}}$ & Interval between edge-flip attempts & Eq. \ref{eq:flip_energy} \\
$P_{\text{flip}}$ & Probability an edge is proposed for a flip & Eq. \ref{eq:flip_energy} \\
$\Delta U_{\text{area}}$, $\Delta U_{\text{tether}}$ & Changes in discrete area and tethering energies during a proposed flip & Eq. \ref{eq:flip_energy} \\
$\Delta U_{\text{flip}}$ & Total energy change of a proposed flip & Eq. \ref{eq:flip_energy} \\
$P_{\text{accept}}$ & Flip acceptance probability & Discretization \\
$\F_{\text{contact},i}$ & Contact force on vertex $i$ & Eq. \ref{eq:F_contact} \\
$\F_i$ & Total deterministic nonviscous mechanical force on vertex $i$ & Eq. \ref{eq:discrete_envelope_force_balance} \\
\multicolumn{3}{l}{\textbf{Steric interactions}} \\
$U_{\text{WCA}}$ & WCA potential & Eq. \ref{eq:wca_potential}, Main Eq. \ref{M-eq:wca_potential} \\
$d$ & Separation distance & Eq. \ref{eq:wca_potential} \\
$\varepsilon$ & WCA interaction strength & Eq. \ref{eq:wca_potential} \\
$\sigma$ & WCA interaction length & Eq. \ref{eq:wca_potential} \\
$U_{\text{ne-spb}}$ & Envelope-complex steric energy & Eq. \ref{eq:U_ne_spb} \\
$\mathcal{V}_{\text{spb}}$, $\x^{\text{spb}}_k$ & Set of SPB-chromatin complex vertices and their positions & Eq. \ref{eq:U_ne_spb} \\
$U_{\text{ne-mt}}$ & Envelope-bundle steric energy & Steric interactions \\
$d_i^{\text{mt}}$ & Minimum distance from envelope vertex $i$ to the bundle surface & Steric interactions \\
\multicolumn{3}{l}{\textbf{Microtubule bundle}} \\
$\X_{\text{mt}}$ & Bundle midpoint position & Fig. \ref{fig:mt_bundle_config_and_velocities} \\
$\U_{\text{mt}}$ & Bundle orientation matrix & Fig. \ref{fig:mt_bundle_config_and_velocities} \\
$\vecU_x$, $\vecU_y$, $\vecU_z$ & Bundle body frame vectors; $\vecU_z$ along the bundle axis & Fig. \ref{fig:mt_bundle_config_and_velocities} \\
$L$ & Total bundle length & Fig. \ref{fig:mt_bundle_config_and_velocities} \\
$L_{\text{half}}$ & Half-bundle length & Eq. \ref{eq:polymerization_rate} \\
$a$ & Half-bundle radius & Bundle \\
$v_{\text{poly}}$ & Microtubule polymerization speed & Eq. \ref{eq:polymerization_rate} \\
$\ell$ & Overlap length $2L_{\text{half}}-L$ & Eq. \ref{eq:overlap_length} \\
$f_{\text{motor}}$ & Extensile motor force per unit overlap length & Bundle \\
$F_{\text{max}}$ & Maximum total extensile force & Bundle \\
$\mu$ & Nucleoplasm viscosity & Bundle \\
$\zeta_{\parallel}$, $\zeta_{\perp}$, $\zeta_r$ & Half-bundle drag coefficients per unit length & Bundle \\
$\boldsymbol{\Gamma}^{\text{mt}}$, $\boldsymbol{\Gamma}_r^{\text{mt}}$ & Translational and rotational drag tensors of the bundle & Eqs. \ref{eq:mt_bundle_translational_drag}, \ref{eq:mt_bundle_rotational_drag} \\
$\Gamma_L$ & Sliding drag & Eq. \ref{eq:mt_bundle_sliding_drag} \\
$\V_{\text{mt}}$, $\vOmega_{\text{mt}}$ & Midpoint velocity and spatial angular velocity & Bundle \\
$v_L$ & Sliding velocity $\dv[L]{t}$ & Bundle \\
$\F^{\text{viscous}}$, $\torque^{\text{viscous}}$, $Q_L^{\text{viscous}}$ & Viscous force, torque, and generalized force & Bundle \\
$\F^{\text{spb}}$, $\torque^{\text{spb}}$, $Q_L^{\text{spb}}$ & Force, torque, and generalized force from the complexes & Bundle \\
$\F^{\text{ne}}$, $\torque^{\text{ne}}$, $Q_L^{\text{ne}}$ & Force, torque, and generalized force from envelope contact & Bundle \\
$\F^{\text{extensile}}$, $\torque^{\text{extensile}}$, $Q_L^{\text{extensile}}$ & Force, torque, and generalized force from motors & Bundle \\
\multicolumn{3}{l}{\textbf{SPB-chromatin complexes}} \\
$r_{\text{complex}}$ & Effective radius of an SPB-chromatin complex & SPB complex \\
$\X_{\text{spb}}$, $\U_{\text{spb}}$ & Position and orientation of a complex & SPB complex \\
$\X_{\text{spb},\pm}$ & Positions of the two SPB-chromatin complexes & Main, spindle positioning \\
$\X_{\text{end}}$ & Microtubule bundle endpoint position & SPB complex \\
$\F^{\text{mt}}$, $\torque^{\text{mt}}$ & Restoring force and torque from the microtubule bundle & Eq. \ref{eq:restoring_force_torque} \\
$\F^{\text{ne}}$, $\torque^{\text{ne}}$ & Force and torque exerted on a complex by the NE & SPB complex \\
$k_{\text{spb}}$, $k_{\text{rot}}$ & Linear and torsional spring constants & Eq. \ref{eq:restoring_force_torque} \\
$\Delta\X_{\text{spb}}$ & Displacement of the complex relative to the bundle endpoint & Eq. \ref{eq:restoring_force_torque} \\
$\Delta\vtheta_{\text{spb}}$ & Rotation vector from bundle frame to complex frame & Eq. \ref{eq:restoring_force_torque} \\
$\boldsymbol{\Gamma}^{\text{spb}}$, $\boldsymbol{\Gamma}_r^{\text{spb}}$ & Translational and rotational drag tensors of the complex & Eq. \ref{eq:spb_chromatin_drag_tensors} \\
\multicolumn{3}{l}{\textbf{Langevin dynamics and integration}} \\
$\X$, $\U$ & Position and orientation of a discrete rigid body & Langevin dynamics \\
$\F$, $\torque$ & Deterministic nonviscous force and torque & Langevin dynamics \\
$\boldsymbol{\Gamma}$, $\boldsymbol{\Gamma}_r$ & Translational and rotational drag tensors & Langevin dynamics \\
$\M$, $\M_r$ & Translational and rotational mobility tensors & Langevin dynamics \\
$\mathbf{W}$, $\mathbf{W}_r$ & Standard Wiener processes & Eqs. \ref{eq:translational_sde}, \ref{eq:rotational_sde} \\
$t_n$, $\Delta t_n$ & Time and timestep at step $n$ & Eq. \ref{eq:discrete_translational_sde} \\
$\Delta t_{\text{max}}$ & Maximum timestep & Main Table \ref{M-tab:model_params} \\
$\Delta\X_n$ & Translational increment & Eq. \ref{eq:discrete_translational_sde} \\
$\veta_n$, $\veta_{r,n}$ & Vectors of independent standard normal variables & Langevin dynamics \\
$\M_r^B$, $\torque^B$ & Body-frame rotational mobility and torque & Eq. \ref{eq:rotational_sde} \\
$\vOmega^B$ & Body-frame angular velocity & Eq. \ref{eq:rotational_sde} \\
$\dd{\vtheta}$ & Differential body-frame angular increment & Eq. \ref{eq:rotational_sde} \\
$\Delta\vtheta_n$ & Finite body-frame angular increment & Langevin dynamics \\
\multicolumn{3}{l}{\textbf{Units and dimensionless groups}} \\
$R$ & Characteristic nuclear radius; unit of length & Main, Methods \\
$R^3\mu/B$ & Unit of time & Main, Methods \\
$\lambda$ & Force ratio $F_{\text{max}}R/B$ & Main Eq. \ref{M-eq:dimensionless_groups} \\
$\rho$ & Size ratio $r_{\text{complex}}/R$ & Main Eq. \ref{M-eq:dimensionless_groups} \\
$\psi$ & Rate ratio $R^2\mu v_{\text{poly}}/B$ & Main Eq. \ref{M-eq:dimensionless_groups} \\
\multicolumn{3}{l}{\textbf{Numerical model validation}} \\
$f_{\text{bend}}$ & Normal component of bending traction & Main, model validation \\
$H^{\text{n}}$, $H^{\text{a}}$ & Numerical and analytical mean curvature & Main, model validation \\
$f_{\text{bend}}^{\text{n}}$, $f_{\text{bend}}^{\text{a}}$ & Numerical and analytical normal bending traction & Main, model validation \\
$\varepsilon_H$ & Dimensionless mean curvature error & Main, model validation \\
$\varepsilon_{\text{bend}}$ & Dimensionless bending traction error & Main, model validation \\
\multicolumn{3}{l}{\textbf{Thermal fluctuation analysis}} \\
$\mathbf{c}$ & NE centroid & Main, fluctuation analysis \\
$R_0$ & Radius of reference sphere & Main, fluctuation analysis \\
$\theta$, $\phi$ & Spherical coordinates & Main, fluctuation analysis \\
$\mathbf{e}_r$ & Radial unit vector & Main, fluctuation analysis \\
$u$ & Dimensionless radial perturbation & Main, fluctuation analysis \\
$\ell$, $m$ & Spherical harmonic degree and order & Main, fluctuation analysis \\
$Y_{\ell m}$ & Spherical harmonic of degree $\ell$ and order $m$ & Main, fluctuation analysis \\
$u_{\ell m}$ & Spherical harmonic expansion coefficient & Main, fluctuation analysis \\
$C_\ell$ & Mean-squared fluctuation spectrum averaged over $m$ & Main, fluctuation analysis \\
$\widehat{C}_\ell$ & Estimated fluctuation spectrum & Main, fluctuation analysis \\
$\ell_{\text{cut}}$ & Maximum spherical harmonic degree included in fit & Main, fluctuation analysis \\
$\gamma_{\text{MS}}$ & Uniform tension in Milner-Safran theory & Main, fluctuation analysis \\
$\Sigma$ & Dimensionless tension $\gamma_{\text{MS}}R_0^2/B$ & Main, fluctuation analysis \\
$B_{\text{fit}}$ & Fitted bending modulus & Main, fluctuation analysis \\
$\Sigma_{\text{fit}}$ & Fitted dimensionless tension & Main, fluctuation analysis \\
$\gamma_{\text{fit}}$ & Fitted uniform tension & Main, fluctuation analysis \\
$N_t$ & Number of sampled configurations & Main, fluctuation analysis \\
$t_k$ & Time of sampled configuration $k$ & Main, fluctuation analysis \\
$\Delta t_{\text{sample}}$ & Sampling interval & Main, fluctuation analysis \\
$\overline{(\cdot)}$ & Surface average & Main, reduced-order analysis \\
$\bar{\gamma}$ & Surface-averaged tension & Main, fluctuation analysis \\
$\langle\bar{\gamma}\rangle$ & Ensemble- and surface-averaged tension & Main, fluctuation analysis \\
\multicolumn{3}{l}{\textbf{Spindle positioning and nuclear shape observables}} \\
$\alpha$ & SPB antipodality & Main, spindle positioning \\
$\r_\pm$ & SPB positions relative to envelope centroid & Main, spindle positioning \\
$A$, $A_0$ & NE area and initial area & Main, area and volume variations \\
$\delta z$ & Dimensionless axial position relative to the spindle midpoint & Main, reduced-order analysis \\
$q_1$ & Second moment of the dimensionless axial position & Main, reduced-order analysis \\
$q_2$ & Normalized third moment of the dimensionless axial position & Main, reduced-order analysis \\
\end{longtable}
}

\section*{Detailed model description}

\subsection*{Nuclear envelope}

We model the nuclear envelope as a two-dimensional fluid membrane whose internal mechanics are dominated by bending elasticity and surface tension. This choice is motivated by the absence of lamins in fission yeast and by previous work showing that many large-scale features of nuclear geometry can be captured by vesicle-like membrane mechanics~\cite{yam11, lim07}. We also consider interactions with the surrounding cyto- and nucleoplasm which we model using an approximate enclosed volume constraint, viscous dissipation, and thermal fluctuations. Finally, steric interactions couple the envelope mechanically to the spindle and chromatin.

Let $M$ denote the NE surface and $\mathbf{r}(s,t)$ its embedding in $\mathbb{R}^3$, with local material coordinates $s=(s^1,s^2)$. We denote the covariant and contravariant tangent bases by $\mathbf{e}_\alpha$ and $\mathbf{e}^\alpha$, respectively, and the outward unit normal by $\mathbf{n}$. The mean and Gaussian curvatures are denoted by $H$ and $K$, with the convention that $H$ is negative for a sphere. We use $\nabla_\alpha$ for the covariant derivative, $\nabla_M$ for the surface gradient, and $\Delta$ for the negative semi-definite Laplace-Beltrami operator~\cite{chavel06}.

Internal envelope forces are described by a surface stress tensor~\cite{sahu17, capovilla17, deserno15}
\begin{gather}
\stensor
=
\stensor^\alpha\otimes\e_\alpha,
\end{gather}
defined such that the force per unit length exerted along the boundary $\partial P$ of a patch $P\subset M$ by the neighboring material outside of $P$ is given by
\begin{gather}
\stensor\cdot\mathbf{u}
=
\stensor^\alpha u_\alpha
\end{gather}
where $\mathbf{u}=u_\alpha\e^\alpha$ is the outward-pointing unit conormal to $\partial P$. We consider two contributions to the surface stress tensor,
\begin{gather}
\stensor
=
\stensor_{\text{bend}}
+
\stensor_{\text{tension}},
\end{gather}
due to bending elasticity $\stensor_{\text{bend}}$ and surface tension $\stensor_{\text{tension}}$.

The energetic cost of bending is described by the energy
\begin{equation}
\label{eq:bending_energy}
U_{\text{bend}} = \int_{M} 2B H^2 \dd{A},
\end{equation}
where $B$ is a bending modulus and we have set spontaneous curvature to zero~\cite{helfrich73}. The associated contribution to the surface stress tensor is given by
\begin{gather}
\stensor_{\text{bend}}
=
2BH
(
\nabla_M\n
+
H\e^\alpha\otimes\e_\alpha
)
-
2B\n\otimes\nabla_M H.
\end{gather}
Here we have neglected saddle-splay contributions to the bending energy because $M$ remains a closed surface with fixed spherical topology throughout our simulations. The saddle-splay term becomes relevant for membranes with changing topologies~\cite{gompper98} and contributes bending moments along membrane boundaries; however, it makes no contribution to the surface stress tensor~\cite{capovilla17, deserno15}.

During mitosis in \pombe, membrane is redistributed from the endoplasmic reticulum to the nuclear envelope, allowing the envelope area to increase. We therefore adopt a constitutive equation for surface tension of the form
\begin{gather}
\gamma(s)=\gamma_0+\gamma_{\text{local}}(s)
\end{gather}
where $\gamma_0$ is a spatially uniform contribution associated with the chemical potential of adding additional lipids to $M$, and $\gamma_{\text{local}}(s)$ is a spatially varying contribution that captures the effect of local deviations in lipid density from equilibrium~\cite{deserno15}. We assume $\gamma_{\text{local}}$ increases linearly with a local measure of area strain which is represented in the discretized model using deviations of individual face areas from a reference value (see Eq.~\ref{eq:face_tension}). Surface tension contributes an isotropic term
\begin{gather}
\stensor_{\text{tension}}
=
\gamma\e^\alpha\otimes\e_\alpha
\end{gather}
to the surface stress tensor.

Because \pombe\ undergoes closed mitosis without global nuclear envelope breakdown, we assume that the enclosed nuclear volume remains approximately conserved over the timescale of the simulations~\cite{lim07,yam11}. Enclosed volume $V$ is regulated by a potential
\begin{gather}
    U_{\text{vol}} = \frac{k_{\text{vol}}}{2V_0} (V - V_0)^2
\end{gather}
where $V_0$ is the reference volume and $k_{\text{vol}}$ is a volume expansion modulus. This results in a traction
\begin{gather}
    -\fdv[U_{\text{vol}}]{\r}=-p\n
\end{gather}
acting normal to the surface where the effective pressure is given by $p=\frac{k_{\text{vol}}}{V_0} (V - V_0)$.

Viscous coupling to the surrounding fluid is modeled with a local drag approximation,
\begin{gather}
\label{eq:f_viscous}
    \f_{\text{viscous}} = -\zeta \vecv,
\end{gather}
where $\zeta$ is an isotropic drag coefficient per unit area and $\vecv$ is the local surface velocity. Thermal fluctuations of the envelope surface are represented by a stochastic traction field $\f_{\text{thermal}}$, which we model as Gaussian white noise with zero mean
\begin{gather}
\label{eq:f_thermal_mean}
\texpect{\f_{\text{thermal}}(s, t)}
=
0
\end{gather}
and covariance prescribed by Eq.~\ref{eq:f_viscous} and the fluctuation-dissipation theorem
\begin{gather}
\label{eq:f_thermal_covariance}
\texpect{\f_{\text{thermal}}(s, t) \f_{\text{thermal}}(s', t')^T}
=
2k_{B}T\zeta\mathbf{I}\delta_M(s,s')\delta(t-t').
\end{gather}
Here $\mathbf{I}$ is the 3-by-3 identity matrix, $\delta_{M}(s,s')$ is the Dirac delta distribution on $M$, and $\delta(t-t')$ is the Dirac delta distribution in time.

Because the covariance in Eq.~\ref{eq:f_thermal_covariance} depends on the instantaneous envelope geometry through $\delta_M$, the stochastic forcing is configuration dependent. Compatibility of the resulting overdamped Langevin dynamics with equilibrium statistical mechanics requires an additional thermal drift contribution~\cite{lau07}. A rigorous continuum treatment requires further choices, including a measure on the space of membrane configurations and additional specification of the stochastic model needed to give meaning to otherwise formal expressions and potentially divergent quantities~\cite{sigurdsson13, cates22, delong14}. We therefore complete the thermodynamically consistent stochastic description after spatial discretization of the envelope, where the thermal drift can be defined in terms of finite-dimensional drag tensors associated with mesh vertices.

Neglecting inertia, local stress balance on the envelope is described by
\begin{gather}
\label{eq:traction_balance}
\zeta\mathbf{v}
=
\nabla_\alpha\stensor_{\text{bend}}^\alpha
+
\nabla_\alpha\stensor_{\text{tension}}^\alpha
-p\n
+
\f_{\text{thermal}}
+
\f_{\text{contact}}
\end{gather}
where the divergences are given by
\begin{gather}
\nabla_\alpha\stensor_{\text{bend}}^\alpha
=
-
2B
\left[
\lbo H
+
2H
\left(
H^2-K
\right)
\right]
\n
\\
\nabla_\alpha\stensor_{\text{tension}}^\alpha
=
\nabla_M \gamma
+
2H\gamma\n
\end{gather}
and $\f_{\text{contact}}$ represents tractions exerted on the envelope by objects like the spindle and chromatin.

\subsubsection*{Spatial discretization}

\begin{figure*}
    \centering
\includegraphics[width=6.75in]{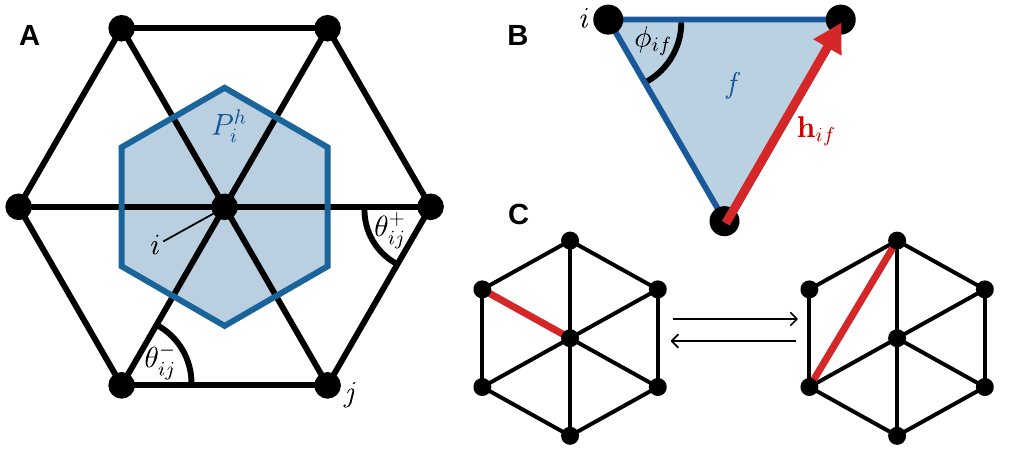}
\caption{\small Discrete surface geometry and mesh connectivity. 
(A) The dual cell $P_i^h$ of vertex $i$ (shown in
  blue) is obtained by connecting triangle centroids with edge
  midpoints. Angles $\theta_{ij}^{\pm}$ opposite edges incident on $i$
  appear in the cotangent formula \ref{eq:cotan_formula} used to
  approximate the Laplace-Beltrami operator. (B) Schematic showing
  notation. (C) Edge flips are used to model membrane fluidity and to
  maintain numerical stability.}
    \label{fig:triangle_mesh_fig}
\end{figure*}

In our simulations, the smooth envelope surface $M$ is approximated by a piecewise-linear triangulated surface $M^h$ characterized by a mesh length scale $h$. We denote the sets of vertices, edges, and triangular faces of $M^h$ by $\mathcal{V}$, $\mathcal{E}$, and $\mathcal{F}$, respectively. Each mesh vertex $i$ is assigned a dual cell $P_{i}^h$ by connecting triangle centroids with edge midpoints to form a closed boundary around the vertex (Fig.~\ref{fig:triangle_mesh_fig}). The dual cell $P_{i}^h$ is then associated with a smooth patch $P_i\subset M$ by mapping each point in $P_{i}^h$ to the closest point in $M$~\cite{hildebrandt06}. To obtain an evolution law for $M^h$, we integrate the continuum traction balance equation~\ref{eq:traction_balance} over each smooth patch $P_i$ and approximate the resulting surface integrals using quantities defined on the corresponding dual cell $P_i^h$.

The viscous drag term is discretized by approximating the average velocity over $P_i$ by the velocity of the associated mesh vertex. With $a_i^*=\text{Area}(P_i^h)$, we take
\begin{gather}
\label{eq:dual_cell_drag}
\zeta\int_{P_i}\mathbf{v}\dd{A}
\approx
\zeta a_i^*\dv[\x_i]{t},
\end{gather}
effectively assigning each vertex an isotropic drag tensor
\begin{gather}
\label{eq:dual_cell_drag_tensor}
\boldsymbol{\Gamma}_i
=
\zeta a_i^*\mathbf{I}
\end{gather}
proportional to its dual cell's area.

The bending contribution
\begin{gather}
\F_{\text{bend}, i}
\approx
-2B\int_{P_i}
\left[\lbo H + 2 H(H^2-K)\right]\n\dd{A}
\end{gather}
requires discrete approximations of the surface normal, mean curvature, Gaussian curvature, and Laplace-Beltrami operator. We obtain an approximate surface normal $\n_i$ at vertex $i$ by averaging the piecewise-constant normal $\n^h$ on $M^h$ over the dual cell,
\begin{gather}
\tilde{\n}_i
=
\frac{1}{a_i^*}\int_{P_i^h}\n^h\dd{A^h},
\end{gather}
and normalizing to get a unit vector
\begin{gather}
\label{eq:vertex_normal}
\n_i=\frac{\tilde{\n}_i}{|\tilde{\n}_i|}.
\end{gather}
For the definition of $P_i^h$ used here (Fig.~\ref{fig:triangle_mesh_fig}), this is equivalent to an area weighted average of normals of faces incident on $i$ followed by normalization.

Mean curvature is approximated by discretizing the identity
\begin{gather}
    \lbo\r=2H\n
\end{gather}
using the cotangent Laplacian~\cite{meyer03, hildebrandt06, wardetzky08}. For a scalar or vector-valued quantity $w$ defined at mesh vertices, we define the cotangent Laplacian by
\begin{gather}
\label{eq:cotan_formula}
    (\cotlap w)_i
    =
    \frac{1}{2 a_{i}^{*}}
    \sum_{j\in \mathcal{V}_i}
    \left(
    \cot\theta_{ij}^{+}+\cot\theta_{ij}^{-}
    \right)
    \left(
    w_j-w_i
    \right),
\end{gather}
where $\mathcal{V}_i$ is the set of vertices adjacent to $i$ and $\theta_{ij}^{\pm}$ are the angles opposite to the edge $(i, j)$ (see Fig.~\ref{fig:triangle_mesh_fig}). When applied to vertex positions $\left\lbrace\x_i|i\in\mathcal{V}\right\rbrace$, the cotangent formula~\ref{eq:cotan_formula} may be interpreted as an approximation to the integral of $\lbo\r$ over $P_i$,
\begin{gather}
\int_{P_i}\lbo\r\dd{A}\approx  a_i^*(\cotlap \x)_i.
\end{gather}
We therefore define the discrete mean curvature at a vertex $i$ by
\begin{gather}
\label{eq:discrete_mean_curvature}
H_i = \frac{1}{2}\n_i\cdot(\cotlap\x)_i
\end{gather}
with $\n_i$ defined by~\ref{eq:vertex_normal}. The Laplace-Beltrami operator acting on mean curvature, $\lbo H$, is likewise approximated by applying the cotangent formula to the nodal values $\left\lbrace H_i|i\in\mathcal{V}\right\rbrace$. Because $H$ is itself obtained from second order derivatives of the embedding function, the subsequent approximation of $\lbo H$ is particularly sensitive to discretization error and mesh irregularity~\cite{guckenberger16, guckenberger17}. This sensitivity is mitigated in part by the mesh-regularization procedures described below, including edge-length tethering and Monte Carlo edge flips. Gaussian curvature is approximated using the angle defect associated with vertex $i$~\cite{meyer03},
\begin{gather}
\label{eq:discrete_gaussian_curvature}
K_i
=
\frac{1}{a_i^*}
\left(
2\pi - \sum_{f\in\mathcal{F}_i}\phi_{if}
\right),
\end{gather}
where $\phi_{if}$ is the interior angle of face $f$ at vertex $i$ and $\mathcal{F}_i$ is the set of faces incident on $i$ (see Fig.~\ref{fig:triangle_mesh_fig}). This definition of $K_i$ is motivated by the fact that the angle defect provides an approximation of the integrated Gaussian curvature
\begin{gather}
\int_{P_i}K\dd{A}
\approx
2\pi - \sum_{f\in\mathcal{F}_i}\phi_{if}
\end{gather}
as a consequence of the Gauss-Bonnet theorem. Combining these approximations gives the expression for the discrete bending force
\begin{gather}
\label{eq:F_bend}
\F_{\text{bend}, i}
=
-2Ba_i^*\left[(\cotlap H)_i + 2 H_i(H_i^2-K_i)\right]\n_i.
\end{gather}

To obtain the surface tension contribution to the force on vertex $i$, we first apply the surface divergence theorem to the corresponding continuum force,
\begin{gather}
\int_{P_i}\nabla_\alpha \stensor_{\text{tension}}^\alpha\dd{A}
=
\oint_{\partial P_i}\gamma \mathbf{u}\dd{s},
\end{gather}
where $\dd{s}$ is the length element along $\partial P_i$ and $\mathbf{u}$ is the outward-pointing unit conormal to $\partial P_i$. We approximate this boundary integral using the corresponding boundary and geometric quantities defined on the triangulated surface. To represent the dependence of surface tension on local area strain, we assign each face $f$ a tension,
\begin{gather}
\label{eq:face_tension}
\gamma_f
=
\gamma_0
+
\frac{k_{\text{area}}}{a_0}(a_f-a_0)
\end{gather}
where $ k_{\text{area}}$ is an area expansion modulus, $a_f$ is the instantaneous area of face $f$, and $a_0$ is a reference area taken to be the average face area at the start of the simulation. The face values $\left\lbrace\gamma_f|f\in\mathcal{F}\right\rbrace$ define a piecewise constant tension field $\gamma^h$ on $M^h$ whose restriction to $f$ is given by
\begin{gather}
\left.\gamma^h\right|_f=\gamma_f.
\end{gather}
We then approximate the continuum boundary force by
\begin{gather}
\F_{\text{tension}, i}
=
\oint_{\partial P_i^h}\gamma^h \mathbf{u}^h\dd{s^h}
\end{gather}
where $\dd{s^h}$ is the length element along $\partial P_i^h$ and $\mathbf{u}^h$ is the outward-pointing unit conormal to $\partial P_i^h$. For the dual cell construction used here (Fig.~\ref{fig:triangle_mesh_fig}), evaluation of this integral gives
\begin{gather}
\label{eq:F_tension}
\F_{\text{tension}, i}
=
\frac{1}{2}
\sum_{f\in\mathcal{F}_i}
\gamma_f\mathbf{h}_{if}\times \n_f
\end{gather}
where $\n_f$ is the unit normal on $f$ and $\mathbf{h}_{if}$ is the directed edge opposite $i$ in $f$ (see Fig.~\ref{fig:triangle_mesh_fig}). We note that \ref{eq:F_tension} is equivalent to the nodal force
\begin{gather}
\F_{\text{tension}, i}
=
-
\pdv[U_{\text{area}}]{\x_i}
\end{gather}
obtained from the discrete surface energy
\begin{gather}
\label{eq:discrete_area_energy}
U_{\text{area}}
=
\gamma_0 A^h
+
\frac{k_{\text{area}}}{2a_0} \sum_{f\in \mathcal{F}}(a_f-a_0)^2
\end{gather}
where $A^h$ is the total area of $M^h$.

The total force due to pressure acting on $P_i$ is
\begin{gather}
-p\int_{P_i}\n\dd{A}.
\end{gather}
We approximate the spatially uniform pressure $p$ using the volume $V^h$ enclosed by the triangulated surface,
\begin{gather}
p^h
=
\frac{k_{\text{vol}}}{V_0}\left(V^h-V_0\right),
\end{gather}
where the reference volume $V_0$ is the initial volume enclosed by $M^h$ at the start of the simulation. Replacing the integrated smooth surface normal over $P_i$ by the corresponding integral of the piecewise-constant normal $\n^h$ over $P_i^h$ then gives
\begin{gather}
\label{eq:F_pressure}
\F_{\text{pressure},i}
=
-\frac{p^h}{3}\sum_{f\in\mathcal{F}_i}
a_f\n_f.
\end{gather}
As was the case with our surface tension discretization, equation~\ref{eq:F_pressure} is equivalent to the derivative of a discrete energy with respect to vertex position. In particular,
\begin{gather}
\F_{\text{pressure},i}
=
-\pdv[U^h_{\text{vol}}]{\x_i}.
\end{gather}
where
\begin{gather}
U^h_{\text{vol}}
=
\frac{k_{\text{vol}}}{2V_0}(V^h-V_0)^2
\end{gather}
is defined entirely in terms of the triangulated surface.

For an arbitrary surface patch $P\subset M$, define the total thermal force acting on $P$ by
\begin{gather}
\F_{P}(t)
=
\int_{P}\f_{\text{thermal}}(s, t)\dd{A}.
\end{gather}
It follows from equations \ref{eq:f_thermal_mean} and \ref{eq:f_thermal_covariance} that $\F_{P}(t)$ is a Gaussian white-noise force with zero mean
\begin{gather}
\texpect{\F_{P}(t)}=0
\end{gather}
and covariance
\begin{gather}
\texpect{\F_{P}(t)\F_{P'}(t')^T}
=
2k_B T\zeta A_{P\cap P'}\mathbf{I}\delta(t-t')
\end{gather}
where $A_{P\cap P'}$ denotes the area of the intersection of $P$ and $P'$. Applying this result to the patches $P_i$ and approximating the area $\text{Area}(P_i)$ by the area of the corresponding dual cells gives the nodal thermal force
\begin{gather}
\label{eq:F_thermal}
\F_{\text{thermal},i}
=
\sqrt{2 k_B T}\boldsymbol{\Gamma}_i^{1/2}\boldsymbol{\xi}_i
\end{gather}
where $\boldsymbol{\Gamma}_i$ is the drag tensor defined by equation \ref{eq:dual_cell_drag_tensor} and $\left\lbrace\boldsymbol{\xi}_i|i\in\mathcal{V}\right\rbrace$ are independent three-dimensional Gaussian white noise processes satisfying
\begin{gather}
\texpect{\boldsymbol{\xi}_i(t)}=0\\
\texpect{\boldsymbol{\xi}_i(t)\boldsymbol{\xi}_j(t')^T}
=
\mathbf{I}\delta_{ij}\delta(t-t').
\end{gather}

Because the dual cell areas may change as the envelope evolves, the nodal drag tensor $\boldsymbol{\Gamma}_i = \zeta a_i^*\mathbf{I}$ is configuration dependent. Thermodynamic consistency therefore requires that we account for thermal drift associated with the random force $\F_{\text{thermal}, i}$~\cite{lau07}. In the It{\^ o} interpretation, this thermal drift appears as an additional term
\begin{gather}
\label{eq:F_drift}
\F_{\text{drift}, i}
=
\frac{k_BT}{3}(\cotlap\x)_i
\end{gather}
in the vertex force balance equation.

Accurate computation of geometric quantities depends on the quality of triangulation $(\mathcal{V}, \mathcal{E}, \mathcal{F})$. Large variations in triangle areas or edge lengths can degrade discrete approximations of surface normals (Eqn.~\ref{eq:vertex_normal}) and curvatures (Eqns.~\ref{eq:discrete_mean_curvature} and \ref{eq:discrete_gaussian_curvature}), with higher order quantities such as $\lbo H$ and the bending force (Eqn.~\ref{eq:F_bend}) being especially sensitive to mesh irregularity~\cite{meyer03, guckenberger16, guckenberger17}. Our model includes several features that help maintain mesh quality during large envelope deformations. The local area-dependent contribution to $\F_{\text{tension}, i}$ penalizes deviations of individual face areas from the reference area $a_0$. We also constrain variations in edge length through a tethering force and employ a Monte Carlo edge-flipping scheme that tends to favor triangulations with more uniform face areas and edge lengths.

To complement the regulation of face areas provided by the local surface-tension contribution, we introduce a tethering potential to penalize edges that deviate too far from a reference edge length $h_0$. An edge of length $b$ is assigned a tethering energy that depends on the dimensionless edge length  $\chi=b/h_0$~\cite{noguchi05},
\begin{gather}
\label{eq:tethering_energy}
U_{\text{tether}}(\chi)
=
\begin{cases}
    k_{\text{tether}}G(\chi, \chi^{\text{rep}}_{\text{onset}}, \chi^{\text{rep}}_{\infty})
    &
    \chi_{\infty}^{\text{rep}}<\chi<\chi^{\text{rep}}_{\text{onset}}
    \\
    0
    &
    \chi_{\text{onset}}^{\text{rep}} \leq \chi \leq \chi^{\text{att}}_{\text{onset}}
    \\
    k_{\text{tether}}G(\chi, \chi_{\text{onset}}^{\text{att}}, \chi_{\infty}^{\text{att}})
    &
    \chi^{\text{att}}_{\text{onset}} < \chi < \chi_{\infty}^{\text{att}}
\end{cases}
\end{gather}
where $k_{\text{tether}}$ is the tether modulus and
\begin{gather}
G(\chi, \chi_{\text{onset}}, \chi_{\infty})
=
\frac{1}{
\left(
1
-
\frac{|\chi-1|}{|\chi_{\infty}-1|}
\right)
}
\exp\left(\frac{-1}{
\left(
\frac{|\chi-1|}{|\chi_{\text{onset}}-1|}
-
1
\right)
}\right).
\end{gather}
The singularities at $\chi=\chi_{\infty}^{\text{rep}}$ and $\chi=\chi_{\infty}^{\text{att}}$ set minimum and maximum edge lengths $h_{\text{min}}=\chi_{\infty}^{\text{rep}}h_0$ and $h_{\text{max}}=\chi_{\infty}^{\text{att}}h_0$, respectively, while $\chi_{\text{onset}}^{\text{rep}}$ and $\chi_{\text{onset}}^{\text{att}}$ determine the onset of repulsive and attractive penalties. The resulting force on vertex $i$ is
\begin{gather}
\label{eq:F_tether}
\F_{\text{tether}, i}
=
\sum_{(i,j)\in\mathcal{E}_i}
U_{\text{tether}}'\left(\chi_{ij}\right)
\frac{\x_j-\x_i}{h_0 |\x_j-\x_i|}
\end{gather}
where $\mathcal{E}_i$ denotes the set of all edges incident on the
vertex $i$ and $\chi_{ij}=|\x_j-\x_i|/h_0$ is the dimensionless length
of the edge $(i,j)$. At the beginning of each simulation timestep,
$h_0$ is updated to the current average edge length and subsequently
held fixed when evaluating tethering forces.

Mesh quality is further regulated by modifying connectivity through
edge flips (Fig.~\ref{fig:triangle_mesh_fig}). Every
$\tau_{\text{flip}}$ units of simulation time, a subset of edges are
selected from $\mathcal{E}$ as candidates for flipping, with each edge
having probability $P_{\text{flip}}$ of being selected. For each
proposed flip, we compute the change
\begin{gather}
\label{eq:flip_energy}
\Delta U_{\text{flip}}
=
\Delta U_{\text{area}}
+
\Delta U_{\text{tether}}
\end{gather}
in the discrete area and tethering energies, \ref{eq:discrete_area_energy} and \ref{eq:tethering_energy}, that would result from the flip. The proposed flip is then accepted with probability
\begin{gather}
P_{\text{accept}}
=
\mathrm{min}\!\left\lbrace
1,
\exp(-\Delta U_{\text{flip}}/k_BT)
\right\rbrace.
\end{gather}
Because $\Delta U_{\text{flip}}$ includes both area and tethering
contributions, this acceptance criterion tends to favor triangulations
with more uniform face areas and edge lengths. In addition to helping
maintain mesh quality, edge flipping provides a method for controlling
the effective two-dimensional viscosity of the envelope by tuning the
parameters $\tau_{\text{flip}}$ and
$P_{\text{flip}}$~\cite{noguchi05}. Decreasing the frequency of flips,
either by lowering $P_{\text{flip}}$ or increasing
$\tau_{\text{flip}}$, inhibits the relaxation of shear
stresses. Increasing the frequency of flips, on the other hand,
decreases viscosity and results in more fluid-like dynamics.

Spindle-generated forces are transmitted to the envelope through short
range steric interactions with the microtubule bundle and
SPB-chromatin complexes. We represent these interactions using the
Weeks–Chandler–Andersen (WCA) potential~\cite{weeks71}
\begin{gather}
\label{eq:wca_potential}
    U_{\text{WCA}}(d)
    =
    \begin{cases}
    4\varepsilon
    \left[
    \left(\frac{\sigma}{d}\right)^{12}
    -
    \left(\frac{\sigma}{d}\right)^6
    \right]
    +
    \varepsilon & d\leq 2^{1/6} \sigma\\
    0 & d>2^{1/6} \sigma
    \end{cases}
\end{gather}
where $d$ is the separation distance, $\varepsilon$ sets the interaction strength, and $\sigma$ sets the interaction length scale. The truncation at $d = 2^{1/6} \sigma$ is chosen to ensure that the interactions are purely repulsive. The envelope and SPB-chromatin complexes, both represented by triangulated surfaces, interact through pairwise repulsion between mesh vertices. If $\mathcal{V}_{\text{spb}}$ denotes the set of vertices belonging to the SPB-chromatin complexes with positions $\left\lbrace \x^{\text{spb}}_k|k\in \mathcal{V}_{\text{spb}}\right\rbrace$, the total energy due to interactions with the complex is
\begin{gather}
\label{eq:U_ne_spb}
U_{\text{ne-spb}}
=
\sum_{i\in \mathcal{V}}
\sum_{k\in \mathcal{V}_{\text{spb}}}
U_{\text{WCA}}(|\x_i-\x^{\text{spb}}_k|).
\end{gather}
Interactions with the microtubule bundle are instead determined by the minimum distance $d^{\text{mt}}_i$ between each envelope vertex $i$ and the cylindrical bundle surface,
\begin{gather}
U_{\text{ne-mt}}
=
\sum_{i\in \mathcal{V}}
U_{\text{WCA}}(d^{\text{mt}}_i).
\end{gather}
The total contact force exerted on envelope vertex $i$ is then
\begin{gather}
\label{eq:F_contact}
\F_{\text{contact}, i}
=
-\pdv{\x_i}
\left(
U_{\text{ne-spb}}
+
U_{\text{ne-mt}}
\right).
\end{gather}

Nodal force balance is described by the overdamped Langevin equation
\begin{gather}
\label{eq:discrete_envelope_force_balance}
\boldsymbol{\Gamma}_i\cdot\mathbf{v}_i
=
\F_{i}
+
\frac{k_BT}{3}(\cotlap\x)_i
+
\sqrt{2k_BT}\boldsymbol{\Gamma}_i^{1/2}\cdot\boldsymbol{\xi}_i
\end{gather}
where the nonthermal deterministic force $\F_i$ includes contributions defined by equations \ref{eq:F_bend}, \ref{eq:F_tension}, \ref{eq:F_pressure}, \ref{eq:F_tether}, and \ref{eq:F_contact}, and the explicit form of the thermal drift term, $k_BT (\cotlap\x)_i/ 3$, assumes the equation is interpreted in the It{\^o} sense. The nodal force balance equation is then discretized in time using the Euler-Maruyama scheme discussed below.

\subsubsection*{Surface drag coefficient}

To obtain a rough estimate for the surface drag coefficient $\zeta$, we equate the total resistance predicted by Eq.~\ref{eq:f_viscous} to Stokes' drag for a sphere with radius equal to the characteristic nuclear radius $R$. For a sphere undergoing rigid translation with velocity $\mathbf{v}$, Eq.~\ref{eq:f_viscous} gives a total force of $-4\pi R^2\zeta\mathbf{v}$. Equating this expression with Stokes' drag law, $\mathbf{F}_{\mathrm{Stokes}}=-6\pi\mu R\mathbf{v}$, gives
\begin{equation}
    \zeta=\frac{3\mu}{2R}.
\end{equation}

\subsection*{Microtubule bundle}

\begin{figure}
    \centering
    \includegraphics[width=6.75in]{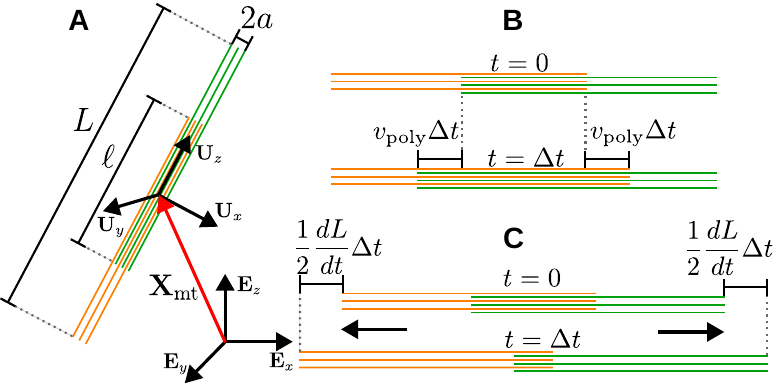}
    \caption{\small (A) The configuration of the microtubule bundle $(\X_{\text{mt}}, \U_{\text{mt}}, L)$ is determined by its midpoint position $\X_{\text{mt}}$, an orthogonal matrix $\U_{\text{mt}}=\vecU_x\E_x^T+\vecU_y\E_y^T+\vecU_z\E_z^T$ representing its orientation, and the total bundle length $L$. (B) Microtubules polymerize at a rate of $v_{\text{poly}}$ which contributes to changes in the overlap length $\ell$. (C) Anti-parallel microtubules slide with respect to one another along the bundle axis, driving changes in both the total bundle length and the overlap length.}
    \label{fig:mt_bundle_config_and_velocities}
\end{figure}

The microtubule bundle (see Figure~\ref{fig:mt_bundle_config_and_velocities}) consists of two partially overlapping antiparallel arrays of microtubules, which we refer to as half-bundles. Each half-bundle is modeled as a cylinder with a fixed radius $a$ and a variable length $L_{\text{half}}$ which grows at a prescribed rate
\begin{gather}
\label{eq:polymerization_rate}
\dv[L_{\text{half}}]{t}=v_{\text{poly}}.
\end{gather}
While the bundle as a whole is free to translate and rotate, internal constraints restrict relative motion of the two half-bundles to sliding along their shared axis. Changes in total bundle length $L$ can occur as the half-bundles slide relative to one another, while changes in the overlap length
\begin{gather}
\label{eq:overlap_length}
\ell = 2L_{\text{half}}-L
\end{gather}
of antiparallel microtubules at the midzone occur as a result of both sliding and microtubule polymerization. Extensile motor forces are represented by a force density $f_{\text{motor}}$ which opposite half-bundles exert on one another in the overlap region. In the simulations described below, half-bundles are assumed to grow at a constant rate until the total extensile force $\ell f_{\text{motor}}$ reaches a maximum value $F_{\text{max}}$, at which point growth ceases.

Due to mechanical constraints between the two half-bundles, the microtubule bundle's independent mechanical degrees of freedom can be characterized by its total length $L$, midpoint position $\X_{\text{mt}}$, and spatial orientation, represented by an orthogonal matrix $\U_{\text{mt}}$, chosen so that $\vecU_z=\U_{\text{mt}}\E_z$ is directed along the bundle axis. Applying d'Alembert's principle to the constrained system gives generalized force balances associated with variations of $\X_{\text{mt}}$, $\U_{\text{mt}}$, and $L$. The first two correspond to the total force and torque balance on the bundle, while the generalized force conjugate to $L$ is determined by the difference between the axial components of forces acting on the two half-bundles. These balances are used to determine the midpoint velocity $\V_{\text{mt}}$, spatial angular velocity $\vOmega_{\text{mt}}$, and sliding velocity $v_L=\dv[L]{t}$, respectively.

Viscous stresses exerted on the half-bundles by the surrounding nucleoplasm are characterized by drag coefficients~\cite{cox70}
\begin{gather}
    \zeta_{\parallel}
    =
    \frac{2\pi\mu}{\ln((L+\ell)/2a)-3/2+\ln(2)}
,\\
    \zeta_{\perp}
    =
    \frac{4\pi\mu}{\ln((L+\ell)/2a)-1/2+\ln(2)}
,\\
\zeta_r = 4\pi\mu a^2,
\end{gather}
where $\mu$ is the nucleoplasm viscosity. The coefficients $\zeta_{\parallel}$, $\zeta_{\perp}$, and $\zeta_{r}$ are associated with translation parallel to the bundle axis, translation perpendicular to the bundle axis, and rotation about the bundle axis, respectively. These local drag coefficients determine effective resistances to midpoint translation, rigid rotation, and relative sliding, given, respectively, by
\begin{gather}
\label{eq:mt_bundle_translational_drag}
\boldsymbol{\Gamma}^{\text{mt}}=\zeta_{\parallel}(L+\ell)\vecU_z\vecU_z^T
+
\zeta_{\perp}(L+\ell)(\mathbf{I}-\vecU_z\vecU_z^T),
\\
\label{eq:mt_bundle_rotational_drag}
\boldsymbol{\Gamma}_r^{\text{mt}}=\zeta_r(L+\ell)\vecU_z\vecU_z^T
+
\frac{1}{12}\zeta_{\perp}
\left(
L^3+\ell^3
\right)(\mathbf{I}-\vecU_z\vecU_z^T),
\\
\label{eq:mt_bundle_sliding_drag}
\Gamma_{L}
=
\frac{1}{4}\zeta_{\parallel}\left(L+\ell\right).
\end{gather}
The corresponding viscous force, torque, and generalized force conjugate to $L$ are $\F^{\text{viscous}}=-\boldsymbol{\Gamma}^{\text{mt}}\cdot\V_{\text{mt}}$, $\torque^{\text{viscous}}=-\boldsymbol{\Gamma}_r^{\text{mt}}\cdot\vOmega_{\text{mt}}$, and $Q_{L}^{\text{viscous}}=-\Gamma_Lv_L$.

Additional generalized forces arise from interactions with the SPB-chromatin complexes, direct steric contact with the nuclear envelope, and extensile forces within the overlap region. We denote the total force and torque exerted on the bundle by the SPB-chromatin complexes as $\F^{\text{spb}}$ and $\torque^{\text{spb}}$, and the corresponding generalized force conjugate to $L$ as $Q_L^{\text{spb}}$. While direct steric interactions with the nuclear envelope produce a force $\F^{\text{ne}}$ and torque $\torque^{\text{ne}}$, they do not contribute to relative sliding. The bundle ends are contained within the SPB-chromatin complexes, so direct envelope-bundle contact occurs only along the curved cylindrical surface, where the contact force is perpendicular to the bundle axis and hence
\begin{gather}
Q_L^{\text{ne}}=0.
\end{gather}
By contrast, the equal and opposite extensile forces exerted between the two half-bundles produce no net force or torque on the bundle,
\begin{gather}
\F^{\text{extensile}}=0, \quad \torque^{\text{extensile}}=0,
\end{gather}
but they produce a generalized force
\begin{gather}
Q_{L}^{\text{extensile}}=\ell f_{\text{motor}}
\end{gather}
that drives relative sliding.

Thermal fluctuations are included in the bundle's midpoint position and orientation dynamics, while relative sliding and polymerization are treated deterministically. Midpoint position and orientation dynamics are governed by the system of equations \ref{eq:translational_sde}, \ref{eq:rotational_sde}, and \ref{eq:frame_reconstruction_sde} with deterministic force and torque
\begin{gather}
\F=\F^{\text{spb}}+\F^{\text{ne}}
,\quad
\torque=\torque^{\text{spb}}+\torque^{\text{ne}}.
\end{gather}
Relative sliding is determined by the balance of generalized forces conjugate to $L$,
\begin{gather}
\Gamma_L v_L = Q_L^{\text{spb}} + Q_L^{\text{extensile}}.
\end{gather}
Finally, as a consequence of Eq.~\ref{eq:polymerization_rate} and the kinematic relation \ref{eq:overlap_length}, the overlap length evolves according to
\begin{gather}
\label{eq:mt_overlap_length_evolution}
\dv[\ell]{t}=2v_{\text{poly}}-v_L.
\end{gather}

\subsection*{SPB-chromatin complexes}

\begin{figure}
    \centering
\includegraphics[width=3.25in]{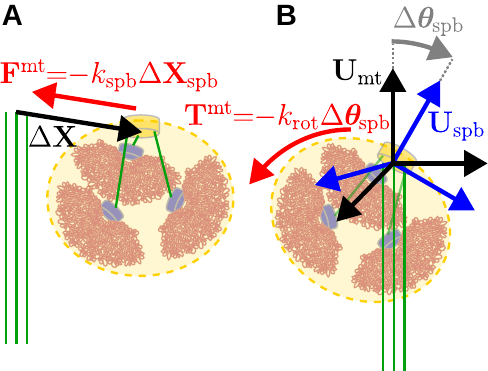}
    \caption{\small The microtubule bundle (green) exerts a restoring force $\F^{\text{mt}}$ (A) and torque $\torque^{\text{mt}}$ (B) on the SPB proportional to the displacement and rotation angle of the SPB relative to the bundle endpoint.}
    \label{fig:mt_spb_interactions}
\end{figure}

At each spindle pole, forces are transmitted from the spindle to the envelope through a collection of objects which are embedded in the envelope or in contact with the inner nuclear membrane. We refer to the collection of objects which are interacting with the envelope at a given time as the SPB-chromatin complex. The characteristic radius $r_{\text{complex}}$ of the complex varies throughout mitosis. During early mitosis $r_{\text{complex}}$ may be close to the radius of a single SPB. However, after chromatin has localized to the spindle poles $r_{\text{complex}}$ may correspond to a sizable fraction of the total nuclear radius $R$. In our simulations, each complex is treated collectively as a single spherical body of radius $r_{\text{complex}}$ whose boundary we represent with a triangulated surface.

Once the contact radius is fixed, the configuration of the SPB-chromatin complex is determined by the position of the SPB $\X_{\text{spb}}$ and an orthogonal matrix $\U_{\text{spb}}$ representing its orientation. As the SPB-chromatin complex translates and rotates relative to the microtubule bundle, it experiences a restoring force and torque,
\begin{gather}
\label{eq:restoring_force_torque}
\F^{\text{mt}}=-k_{\text{spb}}\Delta\X_{\text{spb}}
,\quad
\torque^{\text{mt}}=-k_{\text{rot}}\Delta\vtheta_{\text{spb}},
\end{gather}
that tend to reposition the SPB to coincide with the bundle endpoint and realign the SPB’s material frame with that of the microtubule bundle (Fig.~\ref{fig:mt_spb_interactions}). The restoring force is chosen to be proportional to the difference $\Delta\X_{\text{spb}}=\X_{\text{spb}}-\X_{\text{end}}$ between the SPB position and the endpoint position of the microtubule bundle $\X_{\text{end}}$, while the restoring torque is proportional to the rotation vector $\Delta\vtheta_{\text{spb}}$ that maps the material frame of the microtubule bundle onto the material frame of the SPB.

In addition to the restoring force and torque described above, the complex is subject to steric, viscous, and thermal interactions. Steric repulsion between the complex and nuclear envelope, described by the energy~\ref{eq:U_ne_spb}, results in a net force $\F^{\text{ne}}$ and torque $\torque^{\text{ne}}$ on the complex. Viscous interactions with the nucleoplasm are characterized by translational and rotational drag tensors determined by the complex's spherical geometry
\begin{gather}
\label{eq:spb_chromatin_drag_tensors}
\boldsymbol{\Gamma}^{\text{spb}}=6\pi\mu r_{\text{complex}}\Id
, \quad
\boldsymbol{\Gamma}_r^{\text{spb}}=8\pi\mu r_{\text{complex}}^3\Id.
\end{gather}
Evolution of the SPB-chromatin complex is then governed by the system of equations \ref{eq:translational_sde}, \ref{eq:rotational_sde}, and \ref{eq:frame_reconstruction_sde} using the nonviscous deterministic forces and torques
\begin{gather}
\F=\F^{\text{mt}}+\F^{\text{ne}}, \quad \torque=\torque^{\text{mt}}+\torque^{\text{ne}}.
\end{gather}

\subsection*{Overdamped Langevin dynamics}

Dynamics of discrete elements in our simulations are governed by overdamped Langevin equations, which describe the balance of deterministic force and torque with stochastic thermal fluctuations in the absence of inertia~\cite{lau07, delong14}. All discrete elements have translational degrees of freedom, represented by a position vector $\X$, while the microtubule bundle and SPB-chromatin complexes also have rotational degrees of freedom, represented by an orientation matrix $\U$ that belongs to the special orthogonal group $SO(3)$. For differentiable trajectories, the inertial frame translational velocity $\V$ and angular velocity $\vOmega$ satisfy
\begin{gather}
\dv[\X]{t}=\V
,\quad
\dv[\U]{t}=\hat{\vOmega}\U,
\end{gather}
where the hat operator maps a vector $\mathbf{a}\in\mathbb{R}^3$ to the antisymmetric matrix satisfying $\hat{\mathbf{a}}\mathbf{b}=\mathbf{a}\times\mathbf{b}$ for all $\mathbf{b}\in\mathbb{R}^3$. In general, overdamped translational and rotational motion are coupled through hydrodynamic interactions. For the objects in our model, local hydrodynamic approximations together with body symmetries eliminate viscous translation-rotation coupling resulting in separate translational and rotational drag operators, $\boldsymbol{\Gamma}$ and $\boldsymbol{\Gamma}_r$, which give viscous force and torque
\begin{gather}
\F^{\text{viscous}}
=
-\boldsymbol{\Gamma}\V
,\quad
\torque^{\text{viscous}}
=
-\boldsymbol{\Gamma}_r\vOmega.
\end{gather}
We denote the corresponding translational and rotational mobility operators by $\M=\boldsymbol{\Gamma}^{-1}$ and $\M_r=\boldsymbol{\Gamma}_r^{-1}$. Thermal fluctuations are modeled using Gaussian white noise, with translational and rotational diffusion tensors related to the corresponding mobility operators by the fluctuation-dissipation theorem.

Neglecting inertia, force balance is equivalent to the overdamped It\^o Langevin equation
\begin{gather}
\label{eq:translational_sde}
\dd{\X}
=
\left(
\M\F
+
k_BT\nabla\cdot\M
\right)\dd{t}
+
\sqrt{2k_BT}\M^{1/2} \dd{\mathbf{W}}
\end{gather}
where $\F$ is the deterministic non-viscous mechanical force, $\nabla\cdot\M$ is the divergence of the mobility operator, $\mathbf{W}$ is a three-dimensional standard Wiener process, and the matrix square root is chosen such that $\M^{1/2}\left(\M^{1/2}\right)^T=\M$. In the zero-inertia limit, thermodynamic consistency generally requires a nonzero thermal drift proportional to $\nabla\cdot\M$, as is the case for nuclear-envelope vertices. When $\M$ is independent of position, like for the microtubule bundle and SPB-chromatin complexes, its divergence vanishes. We discretize Eq.~\ref{eq:translational_sde} using the Euler-Maruyama scheme~\cite{higham01},
\begin{gather}
\begin{gathered}
\label{eq:discrete_translational_sde}
\Delta\X_n
=
\left(
\M_n\F_{n}
+
k_BT\nabla\cdot\M_n
\right)
\Delta t_n
+
\sqrt{2k_B T\Delta t_n}\M_{n}^{1/2} \veta_{n},
\\
\X_{n+1}
=
\X_{n}
+
\Delta\X_n
\end{gathered}
\end{gather}
where subscripts denote the timestep at which quantities are evaluated, $\Delta t_n=t_{n+1}-t_n$ is the stepsize, and $\veta_{n}=(\mathbf{W}_{n+1}-\mathbf{W}_{n})/\sqrt{\Delta t_n}$ is a vector of independent standard normal variables. To improve numerical stability, we adapt the timestep based on the maximum deterministic contribution to the velocity of nuclear envelope vertices. The timestep is constrained so that the corresponding displacement $|\M_n\F_n|\Delta t_n$ does not exceed $10\%$ of either the average nuclear envelope edge length or steric interaction cutoff distance. Alternative adaptive timestepping schemes are possible, although care must be taken to avoid biasing the stochastic dynamics~\cite{rackauckas17, sammuller21, leroy24}.

Torque balance is conveniently expressed in the body frame, where the rotational mobility is independent of body orientation for the objects in our model. The body frame mobility and non-viscous mechanical torque are related to their inertial frame counterparts by $\M_r^B=\U^T\M_r\U$ and $\torque^B=\U^T\torque$. Torque balance is equivalent to
\begin{gather}
\label{eq:rotational_sde}
\dd{\vtheta}
=
\M_r^B\torque^B\dd{t}
+
\sqrt{2k_BT} \left(\M_r^B\right)^{1/2} \dd{\mathbf{W}_r}
\end{gather}
where $\dd{\vtheta}$ is a differential angular increment measured in the body frame and $\mathbf{W}_r$ is a three-dimensional standard Wiener process. Because $\M_r^B$ is independent of $\U$, no term analogous to the divergence in Eq. \ref{eq:translational_sde} is required for thermodynamic consistency here. While angular velocity is not defined pointwise when thermal fluctuations are included in the absence of inertia, the differential angular increment can be formally understood as $\dd{\vtheta}=\vOmega^B\dd{t}$ where $\vOmega^B=\U^T\vOmega$ is the body frame angular velocity. The orientation matrix evolves according to
\begin{gather}
\label{eq:frame_reconstruction_sde}
\dd{\U}=\U\circ\dd{\hat{\vtheta}}
\end{gather}
where $\circ$ indicates the equation should be interpreted using Stratonovich conventions. The equivalent It\^o equation includes an additional drift term~\cite{piggott16}. Unlike in the translational case, discretization of rotational dynamics must respect a nonlinear constraint, $\U^T\U=\Id$. To ensure that the orientation matrix remains in $SO(3)$ after each timestep, we discretize the system \ref{eq:rotational_sde} and \ref{eq:frame_reconstruction_sde} using the exponential Euler-Maruyama scheme~\cite{piggott16}. At time $t_n$, we define the finite body frame angular increment
\begin{gather}
\Delta \vtheta_n = \M^{B}_{r,n}\torque^{B}_n\Delta t_n + \sqrt{2k_BT\Delta t_n}\left(\M^{B}_{r,n}\right)^{1/2}\veta_{r, n}
\end{gather}
where $\veta_{r,n}=\left(\mathbf{W}_{r,n+1}-\mathbf{W}_{r, n}\right)/\sqrt{\Delta t_n}$ is a vector of independent standard normal variables, and update body orientation multiplicatively
\begin{gather}
\U_{n+1}
=
\U_n\exp\left(\Delta \hat{\vtheta}_{n}\right)
\end{gather}
using the exponential map for $SO(3)$,
\begin{equation}
    \exp(\hat{\mathbf{a}})
    =
    \mathbf{I}
    +
    \frac{\sin|\mathbf{a}|}{|\mathbf{a}|}
    \hat{\mathbf{a}}
    +
    \frac{1-\cos|\mathbf{a}|}{|\mathbf{a}|^2}
    \hat{\mathbf{a}}^2.
\end{equation}
Related structure preserving integrators use the Cayley transform~\cite{piggott16} or unit quaternions~\cite{delong15, davidchack15} to evolve rigid body orientations. Similar techniques have also been used to model the dynamics of elastic rods and filaments~\cite{lough23a, giusteri18}.

\bibliography{references}